\documentclass{aa}  

\usepackage{graphicx}
\usepackage{txfonts}
\usepackage{amsmath}    
\usepackage{amssymb}    
\usepackage{array,multirow,makecell}
\usepackage{rotating,tabularx}
\usepackage{booktabs, caption, fixltx2e}
\usepackage{threeparttable}
\usepackage{enumitem}
\usepackage{dirtytalk}
\usepackage{float}
\usepackage{placeins}
\usepackage{longtable}
\usepackage{lscape}
\usepackage{adjustbox}
\usepackage{multicol}
\newcommand{\pcm}{\,cm$^{-2}$}  
\newcommand{\msol}{\,M$_\odot$} 
\newcommand{\lsol}{\,L$_\odot$}
\usepackage[colorlinks,linkcolor=blue,citecolor=blue,urlcolor=blue,bookmarks=false,hypertexnames=true]{hyperref}

\begin{document}

   \title{ORion Alma New GEneration Survey (ORANGES) I. Dust continuum and free-free emission of OMC-2/3 filament protostars}
 \titlerunning{ORANGES I. Dust continuum and free-free emission of OMC-2/3 protostars}
   \subtitle{}

   \author{M. Bouvier 
          \inst{1}\fnmsep
          \and
          A. L\'opez-Sepulcre\inst{1,2}
          \and
          C. Ceccarelli\inst{1,3}
          \and
          N. Sakai\inst{4}
          \and
          S. Yamamoto\inst{5,6}
          \and
          Y-L. Yang\inst{4,7}
          }

   \institute{Univ. Grenoble Alpes, CNRS, IPAG, 38000 Grenoble, France\\
              \email{mathilde.bouvier@univ-grenoble-alpes.fr}
         \and
             Institut de Radioastronomie Millim\'etrique (IRAM), 300 rue de la Piscine, 38406 Saint-Martin-D'H\`eres, France
         \and 
            CNRS, IPAG, F-38000 Grenoble, France
        \and
            RIKEN, Cluster for Pioneering Research, 2-1, Hirosawa, Wako-shi, Saitama 351-0198, Japan
        \and 
            Department of Physics, The University of Tokyo, 7-3-1, Hongo, Bunkyo-ku, Tokyo 113-0033, Japan
        \and    
            Research Center for the Early Universe, The University of Tokyo, 7-3-1, Hongo, Bunkyo-ku, Tokyo 113-0033, Japan
        \and 
            Department of Astronomy, University of Virginia, Charlottesville, VA 22904-4235, USA
             }

   \date{Received XXX; accepted YYY}

 
  \abstract
   {The spectral energy distribution (SED) in the millimetre to centimetre range is an extremely useful tool for characterising the dust in protostellar envelopes as well as free-free emission from the protostar and outflow. Actually, the evolutionary status of solar-type protostars is often based on their SED in the near-infrared to millimetre range. In addition, the presence or absence of free-free emission can be considered  an indicator of the source evolutionary stage (Class 0/I versus Class II/III). While many studies have been carried out towards low- and high-mass protostars, little exists so far about solar-type protostars in high-mass star-forming regions, which are likely to be representatives of the conditions where the Solar System was born.}
   {In this work we focus on the embedded solar-type protostars in the Orion Molecular Cloud (OMC) 2 and 3 filaments, which are bounded by nearby HII regions and which are, therefore, potentially affected by the high-UV illumination of the nearby OB stars. We use various dust parameters to understand whether the small-scale structure ($\leq 1000$ au) and the evolutionary status of these solar-type protostars are affected by the nearby HII regions, as  is the case for the large-scale ($\leq 10^4$ au) gas chemical composition. }
   {We used the Atacama Large (sub-)Millimeter Array (ALMA) in the 1.3 mm band (246.2 GHz) to image the continuum of 16 young (Class 0/I) OMC-2/3 solar-type protostars, with an angular resolution of 0.25$''$ (100 au). We completed our data with archival data from the ALMA and VLA VANDAM survey of Orion Protostars at 333 and 32.9 GHz, respectively, to construct the dust SED and extract the dust temperature, the dust emissivity spectral index, the envelope plus disk mass of the sources and to assess whether free-free emission is contaminating their dust SED in the centimetre range.  }
   {From the millimetre to centimetre range dust SED, we found low dust emissivity spectral indexes ($\beta < 1$) for the majority of our source sample and free-free emission towards only 5  of the 16 sample sources. We were  also able to confirm or correct the evolutionary status of the source sample reported in the literature. Finally, we did not find any dependence of the source dust parameters on their location in the OMC-2/3 filament.  }
   {Our results show that the small-scale dust properties of the embedded OMC-2/3 protostars are not affected by the high-UV illumination from the nearby HII regions and that the formation of protostars  likely takes place simultaneously throughout the filament.}


   \maketitle
  
%

\section{Introduction}\label{sec:intro}

Spectral energy distributions (SEDs) from young stellar objects has historically been the main means for classifying their evolutionary status.
Notably, the currently used classification in pre-stellar cores and Class 0 to  II protostars is largely based on the source SED (e.g. \citealt{ LW84,adams1987, lada87, andre93, AM94, evans2009, dunham2014}).
Although studies of gaseous species can provide invaluable informations on the physical, dynamical, and chemical structures of young protostars, the SED remains an important tool for assessing the protostar evolutionary status for several reasons.
First, only the measure of the SED from millimetre (mm) to infrared (IR) wavelengths can provide the luminosity of the young accreting protostar: the smaller the luminosity, the smaller the mass of the central object and its evolutionary age  \citep{andre93, saraceno1996}.
Second, the mm--IR SED can provide the averaged temperature of the dusty envelope plus disk system surrounding the central future star: the lower the temperature the younger the protostar (e.g. \citealt{ML93, chen95, dunham2014}).
Third, the mass of the envelope plus disk system can be estimated by the mm SED once the dust temperature is constrained: the smaller the mass with respect to the guessed central object mass, the more evolved the protostar (e.g. \citealt{andre93, terebey1993, AM94, greene1994}).

Furthermore, dust grains evolve along with the host protostar, both because of the appearance and evolution of icy mantles coating the dust grains (made up of silicates and carbonaceous material: \citealt{hoyle1969, jones2013,jones2017}) and because of  the coagulation process in the cold and dense regions of the protostar (e.g. \citealt{OH94, ormel2009, boogert2015}).
The SED is, again, a major means of studying the grain evolution in protostars and, more relevant to the present work, the coagulation process (e.g. \citealt{MN1993, OH94, draine2006, jones2017}).
Specifically, very often the average grain sizes are estimated by the measurement of the dust emissivity spectral index, $\beta$, in the radio to millimetre wavelength range \citep[e.g.][]{weintraub89, BS91}.

Finally, free-free emission is a mechanism that can dominate the radio emission. It can be extracted from the SED, providing precious information on the evolutionary status of protostars. It probes the ionised gas close to the central object, due to the presence of ionised jets, compact HII regions, or photoevaporating winds, phenomena linked with young protostars (i.e. Class 0 and I sources; e.g. \citealt{andre1988, anglada1995,anglada2018, shirley2007, pascucci2012}. As a consequence, the radio--mm SED is necessary to disentangle the dust from the free-free emission.

In the past such studies have focused on nearby ($\leq 300$ pc) low-mass star-forming regions, both because they are easier to obtain from an observational point of view and because they potentially tell us what could have been the first phases of the Solar System birth \citep[e.g.][]{li2017,murillo2018,hernandezgomez2019}.
As now recognised by several authors, the Solar System was actually born in a region containing numerous low-, intermediate-, and high-mass stars (e.g. \citealt{adams2010, pfalzner2015}). The high-mass stars may have an effect on the physico-chemical conditions of the dusty protostellar envelopes lying nearby, through e.g. higher temperatures, UV radiation, or winds. In particular, several studies have shown an influence of the external UV irradiation from massive stars on the dust temperature of cores or disks \citep{walsh2013, brand2021, haworth2021}.
Studies of solar-mass protostars in this type of star-forming regions are scarcer because they are more distant, and therefore observations are more difficult.

The closest (about 400 pc) low- to high-mass star-forming region is the Orion Molecular Cloud (OMC) complex. In particular, OMC-2 and OMC-3 (also known as OMC-2/3; \citealt{mezger90}) are very active sites of star formation. They are distributed along a filament to the north of the famous BN-KL object. A few studies of OMC-2/3 have appeared in the literature aiming to classify and characterise the population of  Orion protostars, which is the case of the Herschel Orion Protostar Survey (e.g. \citealt{megeath2012, fischer2013, furlan2016}), and to investigate the protostellar disk towards the Class 0 and I protostars of the Orion A and B molecular clouds, which  is the case for The VLA/ALMA Nascent Disk and Multiplicity (VANDAM) Survey of Orion Protostars (e.g. \citealt{tobin2019, tobin2020}).

In this work we focus on the small-scale structure ($\leq$ 1000 au) of more than a dozen embedded protostars in OMC-2/3,  more specifically on their evolutionary status, and we aim to understand whether there is an evolutionary trend along the filament and whether the nearby massive star can affect the dust properties of the protostars. 
To this end, we used new Atacama Large (sub-)Millimetre Array (ALMA) observations, part of our  project ORANGES, and archival data from both ALMA and the Very Large Array (VLA) to obtain the dust SED and basic properties of each source: multiplicity, constraints on the dust temperature at 100 $-$ 1000 au, dust emissivity spectral index $\beta$, free-free emission. Putting together these properties, we were able to constrain the evolutionary status of the targeted protostars and whether the position along the OMC-2/3 filament impacts it.

This paper is organised as follows. 
We first report a literature review of the dust temperature, dust emissivity spectral index $\beta$,  envelope plus disk (envelope + disk) mass, and  free-free emission as a function of the evolutionary status in Sect. \ref{sec:review}.
In Sect. \ref{sec:sources} we describe the ORANGES survey and the selected source sample, and we report a review of what is known about them in terms of dust properties and what is still missing from these previous studies.
The  observations and archival data used are presented in Sect. \ref{sec:observations}. The results of the new ALMA observations, with the source maps, and the source sizes and flux densities extracted, are presented in Sect. \ref{sec:results}. 
The source multiplicity and properties derived by the dust SED analysis (dust temperature, dust emissivity spectral index, and masses), as well as the construction of the dust SEDs, are reported in Sect. \ref{sec:properties}.  Finally we discuss the results in Sect. \ref{sec:discussion} before concluding in Sect. \ref{sec:conclusion}.

\section{Review of protostar properties versus evolutionary status}\label{sec:review}

Several parameters have an impact on the shape of SEDs, such as the dust temperature and emissivity spectral index across the protostar, the (envelope + disk) mass of the source and the presence of free-free emission. We present here the parameters we   focus  on in this work and how they evolve with the different evolutionary stages.

\paragraph{Dust temperature:}
The dust temperature, $T_{\text{d}}$, has an impact on the shape of the SED by setting the position of its peak. We can use the dust temperature to distinguish between pre-stellar cores ($T_{\text{d}}\leq$10 K) and more evolved stages (i.e. Class 0/I/II,  10 $\leq T_{\text{d}}\leq$100 K; e.g. \citealt{ceccarelli2000a,jorgensen2002, bergin2007,dullemond2010,crimier2010}); however, it is hard to distinguish between each of the protostellar stages (Classes 0, I, and II). In that case the bolometric temperature, $T_{\text{bol}}$, defined as the temperature of a black body with the same mean frequency as the observed SED \citep{ML93, andre2000} is more useful. \citet{chen95,chen97}, among others, showed that young stellar objects (YSOs) have  different bolometric temperatures depending on their stages. Roughly speaking, Class 0 sources have $T_{\text{bol}}< 70 $ K, Class I sources have 70 $<T_{\text{bol}}< 650 $ K, and Class II and III sources have $T_{\text{bol}}>650 $ K.

\paragraph{Dust emissivity spectral index:}

The dust emissivity spectral index, $\beta$, is useful to probe dust properties such as grain size, composition, and shape (e.g. \citealt{draine2006, compiegne2011, jones2017}). While a fixed value of $\beta=2$  is commonly adopted in the literature \citep{hildebrand83,DL84}, a change in the composition of the dust grains, large optical depths, or grain growth are usually responsible for lower values of $\beta$ (e.g. \citealt{BS91, OH94, draine2006, guilloteau2011, testi2014, ysard2019}). On the other hand, the growth of the icy grain  mantle can explain higher values ($\beta \geq 2$; e.g. \citealt{kuan1996,lismenten1998}). In  this context, the dust emissivity spectral index can change depending on the evolutionary stage of the sources: The highest values have been measured towards pre-stellar cores (1$ \leq\beta \leq$ 2.7; e.g. \citealt{lis1998, shirley2005,shirley2011, schnee2010, sadavoy2013, bracco2017}),  while the lowest values have been measured towards Class II protostars ($0 \leq\beta\leq 1$; e.g. \citealt{ricci2010a, bracco2017,tazzari2020}). Class 0 and I protostars show intermediate values ($\beta \sim 1.7 - 2$; e.g. \citealt{natta2007}).

\paragraph{Dust mass:}
In young protostars, the envelope mass dominates over  that of the disk. Considering that our observations are not sensitive enough to distinguish between the two components, we focus here only on how the envelope mass evolves with the evolutionary stage of protostars. 
First during the Class 0 phase, the envelope mass dominates the object mass ($M_{\text{env}}\geq M_{\text{*}}$), whilst for Class I sources the envelope mass is lower ($M_{\text{env}}\leq M_{\text{*}}$ and $M_{\text{env}} \geq 0.1$\ \msol; \citealt{andre93, enoch2009}). 
Then, in Class II and III objects the envelope has almost entirely dissipated ($M_{\text{env}}\leq 0.1$\msol; \citealt{greene1994, crapsi2008}) and the mass is dominated by that of the central object.

\paragraph{Free-free emission:}
Finally, the presence of thermal (free-free) emission results in a shallower slope of the SED in the millimetre to centimetre domain and can be used to determine the evolutionary status of the object. This free-free emission is thought to be produced by jets or winds associated with the outflows driven by Class 0 and I sources \citep{anglada1996,pech2016, tychoniec2018a} or could be associated with the  photoevaporation region of Class I disks (e.g. \citealt{pascucci2012, anglada2018}).  At later stages (Class II/III sources), radio continuum emission due to gyrosynchrotron radiation can be detected, but is characterised by a different spectral index \citep{FM85, andre1988,pech2016}. Finally, the absence of such compact radio emission in a source indicates that the latter is likely a pre-stellar core \citep{yun96}.

\section{The  ORANGES project}\label{sec:sources}

\subsection{Description of the project}\label{subsec:oranges-description}
The Orion ALMA New Generation Survey (ORANGES) is a project aiming to study the molecular content, on small scales ($\leq 1000$ au), of the solar-type protostars located in the OMC-2/3 filament.
It follows a large-scale ($\leq 10^4$ au) study of these same protostellar cores \citep{bouvier2020}. 
The focus of this previous study, carried out with the single-dish telescopes IRAM-30m\footnote{\url{https://www.iram-institute.org/}} and Nobeyama-45m\footnote{\url{ https://www.nro.nao.ac.jp/~nro45mrt/html/index-e.html}}, was to identify Warm Carbon Chain Chemistry (WCCC) protostars and hot corino candidates using observations of two species, CCH and CH$_3$OH. 
Briefly, these two molecules are expected to be abundant in WCCC protostars and hot corinos, respectively. 
Therefore, the use of the relative abundance of these two species is thought to be a good indicator of the chemical nature of solar-type protostars \citep[e.g.][]{higuchi2018}. 
However, given the relatively large spatial scales probed by the single-dish observations ($\sim 10^4$ au) and the highly UV illuminated OMC-2/3 region, we found that the line emission from the two species is dominated by the photodissociation region (PDR), and not by the protostars.  
Therefore, in order to assess the chemical nature of the OMC-2/3 solar-type protostars, small-scale ($\leq 1000$ au) interferometric observations are mandatory to get rid of the PDR contamination.

The  ORANGES project consists of new ALMA observations at 246.2 GHz towards the same sample of sources studied in \cite{bouvier2020}. 
The spatial resolution is about 0.$''$25 equivalent to about 100 au at the OMC-2/3 distance.
As said, our ultimate goal is to assess the number of WCCC sources versus hot corinos in the OMC-2/3 region, which is the best known analogue of the Solar System progenitor. 
To achieve this goal we designed the spectral setup to make the distinction between hot corinos and WCCC objects, which will be presented in a forthcoming paper on the molecular content.

In the following we briefly present the OMC-2/3 region and then introduce the ORANGES source sample (\S ~\ref{subsec:source-sample}).
We then review the previous continuum studies obtained towards the sample sources (\S ~\ref{subsec:prev_studies}), and finally we summarise what is still poorly or totally unknown of them with respect to the dust (\S ~\ref{subsec:missing-info}). 

\subsection{Source sample}\label{subsec:source-sample}
The OMC-2/3 filament is located at a distance of (393 $\pm$ 25) pc from the Sun \citep{grosschedl2018} and it is bounded by three HII regions: NGC1977 to the north, the Trapezium OB cluster to the south, and M43 to the southeast. 
It is the closest analogue to our Sun's birth environment, hence the high interest to study this region. 
OMC-2/3 is one of the most active star-forming regions (e.g. \citealt{rayner89, bally1991, jones1994, AD95}) in which a wealth of low-mass protostars have been detected (e.g.  \citealt{mezger90, chini1997, lis1998, nielbock2003}). 
From these studies we previously selected nine bona fide solar-type protostars to study their molecular content \citep[see above;][]{bouvier2020}. 
We applied three criteria to select the sources: (1) detection in the (sub-)mm continuum emission; (2) estimated (from the large-scale continuum) envelope masses $\leq $12\ \msol; (3) bona fide Class 0 and I protostars, based on their estimated bolometric temperature. In ORANGES we targeted the nine sources, centring the observations on the protostar coordinates (Table \ref{tab:beam_param}) derived from previous single-dish observations \citep{chini1997, lis1998,  nielbock2003}.

\subsection{Previous continuum studies on the sample sources}\label{subsec:prev_studies}

All the properties previously known and described below are summarised in Table \ref{tab:new_sources} (see also Sect. \ref{sec:properties}).

\paragraph{Source classification:}
The nine selected sources are also part of the HOPS survey \citep[e.g.][]{megeath2012, fischer2013, furlan2016}, which used \textit{Herschel}\footnote{\url{https://sci.esa.int/web/herschel}} observations.
We therefore already have some information on those protostars, even though the Herschel observations can not disentangle multiple sources in the \textit{Herschel} beam ($5.2 - 12''$).
HOPS provided estimates of the sources evolutionary classification based on their bolometric luminosity, $L_{\text{bol}}$; bolometric temperature, $T_{\text{bol}}$; and envelope masses, $M_{\text{env}}$. 
From the HOPS survey, six of our nine sources were classified as Class 0, two as Class I, and one as a Flat spectrum source. 
The derived bolometric luminosities and temperatures range from 5.7\lsol \ to 23.2\lsol \ and from 28.4 K to 186.3 K, respectively, and the envelope masses range from $\sim 0.1$\msol \ to $\sim 5.6$\msol.   

\paragraph{Dust temperature:}
Large-scale ($>1000 $ au) studies were performed to derive the dust temperature and dust emissivity spectral index, and to determine whether free-free emission is present towards those sources. 
First, the previously derived dust temperature from continuum emission along the filament is typically about 20-30 K \citep{chini1997, lis1998,JB1999,schuller2021}, and similar values were found for the kinetic temperature using molecular lines (e.g. \citealt{li2013, hacar2018}), although \citet{shimajiri2009} found a warmer temperature up to 50 K towards the FIR6 region. 

\paragraph{Dust emissivity spectral index:}
Several studies, covering wavelengths between 350 $\mu$m and 2 mm, found values for the dust emissivity spectral index ranging from 1 to 2 \citep{chini1997, lis1998, JB1999}. Later, \citet{schnee2014} measured the dust emissivity spectral index of the OMC-2/3 sources between 1.2 and 3.3 mm.
They found a median value for the index of 0.9 on 0.1 pc scales, lower than what was previously measured. On the other hand, \citet{sadavoy2016} used \textit{Herschel} data between 160 and 550 $\mu$m with their IRAM-30m data at 2 mm (150 GHz) and derived indexes around 1.6-1.8. Recently, \citet{mason2020} found that the discrepancy between the values from \citet{schnee2014} and \citet{sadavoy2016} was due to abnormally high 3.3 mm (90.9 GHz) emission in OMC-2/3. The reason of the excess emission at 90.9 GHz is not entirely clear, although the possibility of the presence of amorphous dust grains could be an explanation (e.g. \citealt{meny2007, coupeaud2011, paradis2011}). 

\paragraph{Free-free emission:}
Finally, using VLA, \citet{reipurth1999} detected 8.3 GHz emission towards 11 sources associated with the OMC-2/3 filament, among which there are three of our protostars: MMS2, MMS9, and FIR1a. 
They attributed the 8.3 GHz free-free emission as probably due to shocks caused by the out-flowing material.

\subsection{Limits of the previous studies and novelty of ORANGES}\label{subsec:missing-info}
\citet{tobin2020} published high-resolution continuum observations towards our source sample as part of the VANDAM survey. 
The two surveys, VANDAM and ORANGES, are complementary;  the VANDAM survey aims to characterise the dust physical properties of the protostellar disks of the Orion protostars (masses, radii), whereas ORANGES focuses on the dust physical properties of the protostars (inner envelope+disk), specifically of the sources on a larger scale.
The different objectives of the two surveys make them highly complementary, and also means that the ORANGES survey provides novel important information.
We describe the specific differences and the  ORANGES novelties below.

First, in order to derive the disk masses, \citet{tobin2020} used the bolometric luminosity of each system to derive the dust temperature. However, several of our sources being multiple systems, the bolometric luminosity is not accurate for each of the objects in each system. 
We therefore aim to constrain the dust temperature for each of our sources without using bolometric luminosities. 

Second, based on the study from \citet{tychoniec2018b} in the Perseus region, they assumed the 32.9 GHz continuum emission to be dominated by the dust emission even though they mention the possibility of free-free contamination at this wavelength. 
Therefore, the spectral indexes they provide between 333 and 32.9 GHz might not be accurate if the 32.9 GHz emission is contaminated by free-free emission. 
We thus aim to provide a more reliable dust emissivity spectral index for each of our sources, and use these indexes to confirm or correct the evolutionary stages of the sources. 

Finally, \citet{tobin2020} provide the disk masses assuming optically thin dust emission at 333 GHz. 
Therefore, the derived disk masses can be considered only as lower limits if the assumption does not hold and we aim to better constrain the (envelope + disk) masses of our source sample. 

In order to better constrain the physical properties of the ORANGES sample, we chose to complete our 246.2 GHz ALMA data with the archived data from the VANDAM survey \citep{tobin2019a,tobin2019b}. 
The different sets of observations are described in the next section.

\section{Observations and data reduction}\label{sec:observations}

\subsection{ALMA observations at 246.2 GHz}

The ALMA ORANGES observations were performed on October 27--29, 2016, during Cycle 4, in Band 6 ($\sim 250$ GHz), under the ALMA project \textsc{2016.1.00376.S}. A total of 42 antennas of the 12m array were used with a baseline length range of 18.6m - 1100m. The integration time on source is $\sim$ 19 min for each source. As the present study concerns the dust continuum analysis, we present only the setup of the continuum spectral window. With a total bandwidth of $\sim 2$ GHz, the continuum spectral window was centred at 246.2 GHz and with a spectral resolution of 0.977 MHz ($\sim$ 1.2 km/s). The band-pass and flux calibrators used were J0510+1800 and J0522-3627, and the phase calibrators used were J0607-0834 and J0501-0159. The flux calibration error is estimated to be better than 10\%. The precipitable water vapour (PWV) was typically less than 1 mm, the system temperatures less than 200 K, and the phase root mean square (rms) less than 50$^\circ$.

We used the Common Astronomy Software Application (CASA; \citealt{mccmullin2007}) for the data calibration. We then exported the calibrated visibility tables to the GILDAS \footnote{\url{http://www.iram.fr/IRAMFR/GILDAS}} format, and performed the imaging in MAPPING. We produced continuum images by averaging line-free channels in the visibility plane. We produced the continuum images using natural weighting (via the CLEAN procedure). Phase self-calibration was performed for all sources with a signal-to-noise ratio (S/N) $>100$. Only CSO3 did not fulfil the criteria. We applied three successive rounds of calibration with solution intervals in the  range 300-60s. Solution intervals were adjusted from one source to another, depending on the S/N and of the number of flagged solutions. This resulted in a significant improvement in the S/N for the continuum images, from 24\% (FIR1a) to 293\% (SIMBA-a). The maps shown in this paper are not corrected for the primary beam, but we took into account the correction to measure the flux densities. The resulting synthesised beam for each field,  as well as the noise rms, are presented in Table \ref{tab:beam_param}. The half power primary beam size is 25.6$''$. 

\begin{table}[ht]
    \centering
       \caption{Continuum beam parameters at 246.2 GHz, 333 GHz, and 32.9 GHz, as well as rms and centre of observations used. Information on the beam size and rms at 333 and 32.9 GHz are issued from the archived images used \citep{tobin2019a,tobin2019b}, whilst information on the centre of observations for the two sets of data are   from \citet{tobin2020}.}
         \label{tab:beam_param}
    \resizebox{\linewidth}{!}{
    \begin{tabular}{lccccc}
    \hline \hline
         Field& Beam size  & PA  & rms &R.A. &Dec. \\
         &[$''$]&[$^{\circ}$]&[$\mu$Jy/beam]&[J2000] & [J2000]\\
         \hline
          \multicolumn{6}{c}{ALMA 246.2 GHz}\\
          \hline
         CSO33&0.32 $\times$ 0.27 & -77 & 44&05:35:19.5& $-$05:15:35.0\\
         FIR6c &0.32 $\times$ 0.28 & 90 & 60& 05:35:21.6& $-$05:13:14.0\\
         FIR2 &0.32 $\times$ 0.27 & -76 & 54 & 05:35:24.4& $-$05:08:34.0\\
         FIR1a &0.32 $\times$ 0.27 & -76 &58& 05:35:24.4& $-$05:07:53.0\\
         MMS9&0.32 $\times$ 0.27 & -74 & 50 & 05:35:26.2& $-$05:05:44.0 \\
         MMS5&0.32 $\times$ 0.28 & -76 & 80& 05:35:22.5& $-$05:01:15.0\\
         MMS2&0.32 $\times$ 0.27 & -77 & 60& 05:35:18.5& $-$05:00:30.0 \\
         CSO3 &0.32 $\times$ 0.27 & -74 & 49& 05:35:15.8& $-$04:59:59.0\\
         SIMBA-a &0.32 $\times$ 0.27 & -77 & 50&05:35:29.8& $-$04:58:47.0\\
         \hline
          \multicolumn{6}{c}{ALMA 333 GHz}\\
           \hline
        CSO33 & 0.12 $\times$ 0.12 & 29&580 &05:35:19.465& $-$05:15:32.72\\
         FIR6c& 0.13 $\times$ 0.12 &86&710&05:35:21.400&$-$05:13:17.50\\
         FIR2 & 0.12 $\times$ 0.12 &26&277&05:35:24.306&$-$05:08:30.58\\
         FIR1a-a &0.12 $\times$ 0.12&25&518 &05:35:25.607&$-$05:07:57.32\\
         FIR1a-b &0.13 $\times$ 0.12 &84&640&05:35:23.927 & $-$05:07:53.47 \\
         MMS9 & 0.12 $\times$ 0.12 &23&590&05:35:25.824&$-$-05:05:43.65\\
         MMS5 &0.12 $\times$ 0.12 &21& 335&05:35:22.430&$-$05:01:14.15\\
         MMS2 & 0.12 $\times$ 0.12 &17&296&05:35:18.317&$-$05:00:32.97\\
         SIMBA-a &0.12 $\times$ 0.12 &13&280 &05:35:29.720&$-$04:58:48.79\\
         \hline
         \multicolumn{6}{c}{VLA 32.9 GHz}\\
         \hline
         CSO33 & 0.11 $\times$ 0.08&-31&11&05:35:19.466&$-$05:15:32.72 \\
         FIR6c&0.10 $\times$ 0.07 &-15& 17&05:35:21.400&$-$05:13:17.50\\
         FIR2 &0.09 $\times$ 0.07&31&8& 05:35:24.299&$-$05:08:30.73\\
         FIR1a&0.09 $\times$ 0.06 &14&11 &05:35:23.926 & $-$05:07:53.47 \\
         MMS9 & 0.12 $\times$ 0.07&35& 8&05:35:25.823&$-$-05:05:43.65\\
         MMS5 & 0.11 $\times$ 0.06&0&11 &05:35:22.891&$-$05:01:24.21\\
         MMS2 & 0.11 $\times$ 0.08&-33&10&05:35:18.915&$-$05:00:50.86 \\
         CSO3 & 0.12 $\times$ 0.07&42& 14&05:35:16.152&$-$05:00:02.26\\
         SIMBA-a & 0.09 $\times$ 0.06&-1&8 &05:35:29.720&$-$04:58:48.79\\
         \hline
    \end{tabular}
    }
\end{table}

\subsection{Archival data at 333 and 32.9 GHz}
Our source sample is part of the VLA/ALMA Nascent Disk and Multiplicity (VANDAM) Survey of Orion Protostars. 
In order to be able to construct the dust SED of the sources in the centimetre to millimetre range, we used archival data from the VANDAM Survey \citep{tobin2020} at 333 GHz (0.87mm) from ALMA, and we used the data at 32.9 GHz (9mm) from the VLA observations. 

We used the \textit{robust}=2 (equivalent to natural weighting) cleaned images, for both the ALMA and the VLA data. Both sets of  archival data are taken from the Harvard Dataverse\footnote{\url{https://dataverse.harvard.edu}} \citep{tobin2019a, tobin2019b}. The angular resolutions are 0.1$''$ ($\sim 39$ au) and 0.08$''$ ($\sim 31$ au) at 333 GHz and at 32.9 GHz respectively. The baseline length ranges are 15 $-$ 3700m for ALMA and 0.68 $-$ 36.4 km for VLA (A-configuration). The resulting synthesised beam for each field and each interferometer, as well as the rms noise, are presented in Table \ref{tab:beam_param}. More observational details are given in \citet{tobin2020}.


\section{Results} \label{sec:results}

\subsection{Maps}\label{subsec:maps}
We observed nine fields centred on the nine protostars of Table \ref{tab:beam_param}. Thanks to the achieved angular resolution of 0.25$''$ ($\sim 100$ au), many multiple sources could be disentangled. The complete list of detected sources in each field is listed in Table \ref{tab:all_sources}. In total, we detected 28 sources, of which 18   are located inside the primary beam (see Fig. \ref{fig:Maps_large}). 
The following analysis   focuses on the latter, with the exception of FIR6c-c and MIR4. 
For FIR6c-c, no clear counterparts were found in the 333 and 32.9 GHz data. For MIR4, the source is not included in the 333 GHz maps and we did not find any counterpart at 32.9 GHz. 
Therefore, we analysed 16 protostars in this work. Their coordinates and previously attributed parameters are listed in Table \ref{tab:new_sources}. 

\begin{figure*}[ht]
   \centering
    \includegraphics[width=1\linewidth]{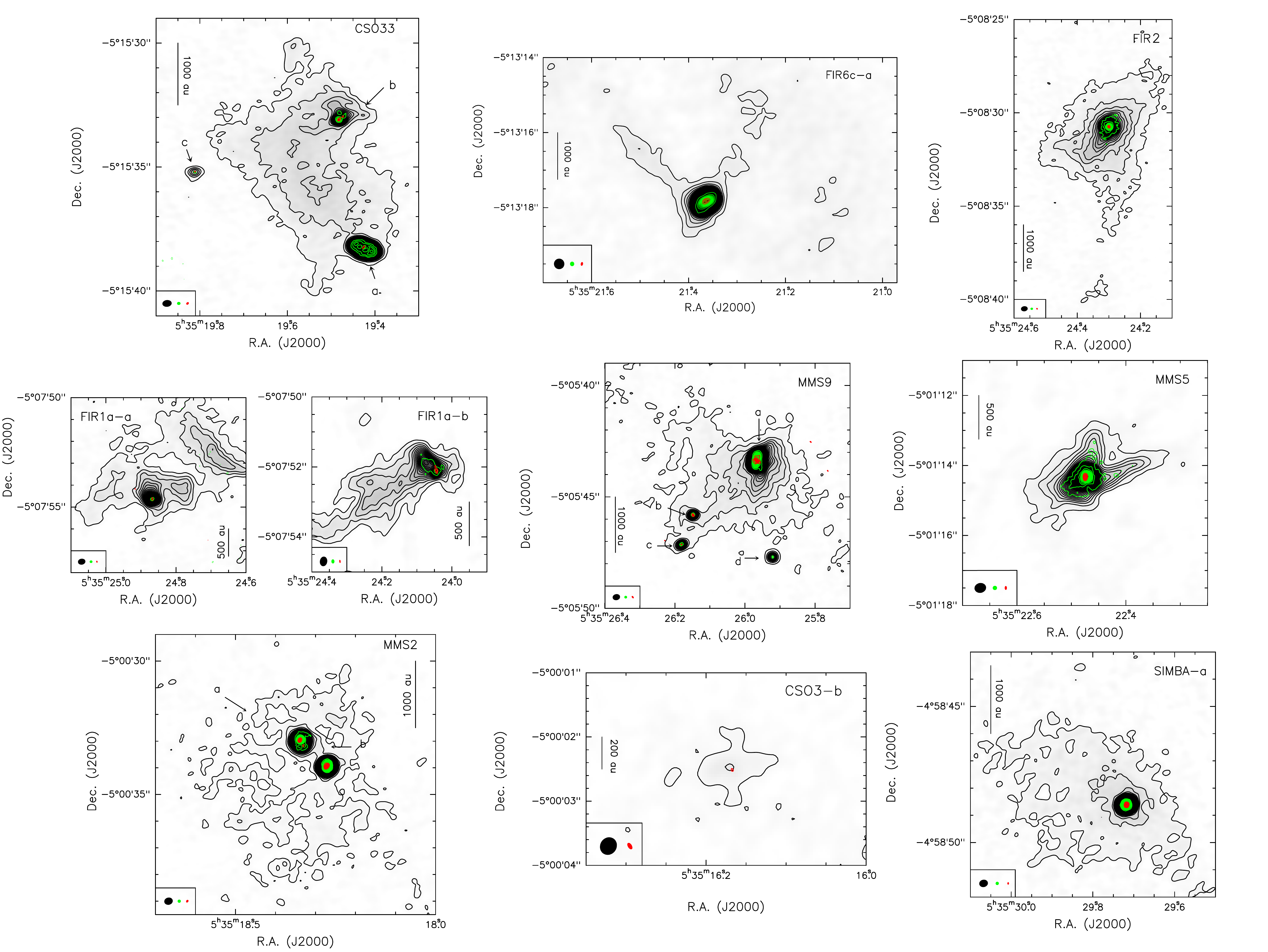}
   \caption{Continuum maps of the nine fields in Table \ref{tab:beam_param}, resolved into several cores, listed in Table \ref{tab:new_sources}. Each panel shows the superposition of the ORANGES 246.2 GHz dust continuum (ALMA, grey shaded area + black contours) with 333 GHz continuum (ALMA, green contours) and 32.9 GHz continuum (VLA, red contours), the latter retrieved from the ALMA archive.
   For the continuum at 246.2 GHz, levels start from 3$\sigma$ with steps of 5$\sigma$, except for FIR6c for which levels starts at 5$\sigma$ and for CSO33, MMS9 and CSO3 for which level steps are of 7$\sigma$. Contours for the 333 GHz data start at 5$\sigma$ with steps of 5$\sigma$, except for CSO33 and FIR1a-a for which contours start respectively at 3$\sigma$ with steps of 3$\sigma$ and at 10$\sigma$ with steps of 5$\sigma$. Contours for the 32.9 GHz data start at 5$\sigma$ with steps of 5$\sigma$, except for FIR1a-a, FIR1a-b, and CSO3-b for which contours start at 3$\sigma$ with steps of 5$\sigma$ and for MMS9 for which contours start at 4$\sigma$ with steps of 5$\sigma$. 
   Table \ref{tab:beam_param} reports the 1$\sigma$ values at each wavelengths. The synthesised beams at 246.2 GHz (black), 333 GHz (green), and 32.9 GHz (red) are depicted in the lower left corner of each panel.  }
    \label{fig:cont_OMC23}
\end{figure*}

The ORANGES 246.2 GHz continuum maps of the 16 sources are shown in Fig. \ref{fig:cont_OMC23}, where the archival 333 GHz ALMA and 32.9 GHz VLA data are superposed. Most of the sources that appeared to be single sources with single-dish observations are multiple systems. Some sources are separated by about 1000 to $\sim$5500 au in the plane of the sky (e.g. CSO33, FIR6c, FIR1a, MMS9); others have smaller projected separations of less than $\sim$700 au, (MMS2-a with MMS2-b, and MMS9-b with MMS9-c) or even less than $\sim 100$ au (CSO33-b and MMS2-a).
Finally, one source, CSO3-b, is part of a binary system, whose other component, CSO3-a, is about 7000 au away and out of the primary beam (see Fig. \ref{fig:Maps_large}). 

All the sources are detected at a $\geq10\sigma$ level. We resolve all the sources at 246.2 GHz, except CSO33-c and MMS9-c, which we only marginally resolve. All of the sources are clearly centrally peaked, except CSO3-b for which the continuum is more diffuse. 
For most of the sources the dust emission at 246.2 GHz always extends  asymmetrically, and in some cases neighbouring sources are `{linked}' by the dust emission (CSO33-a with CSO33-b, MMS9-a with MMS9-b and MMS9-c, and MMS2-a with MMS2-b). 
MMS9-d and CSO33-c are the only two sources of the sample that are compact with no dust emission extension around them. Some sources are clearly elongated at 246.2 GHz, such as FIR6c-a,  CSO33-a, MMS9-a, and FIR1a-b. For the first two sources, this is due to the fact that we resolve the protostellar disk emission \citep{tobin2020, sheehan2020}. The protostellar disk of MMS9-a is also resolved, but it is more obvious when looking at the 333 GHz emission. CSO3-b is the only source for which the 333 GHz emission is not detected, whilst MMS9-d is the only source for which the 32.9 GHz emission is not detected. 

Finally, the 333 and 32.9 GHz emission are always more compact than the 246.2 GHz emission. This may be due to several factors. The first factor would be that VLA observations are more sensitive to larger grains than ALMA observations, therefore tracing a slightly different region than with ALMA observations. Another factor could be the difference in sensitivity: VANDAM ALMA data at 333 GHz are ten times less sensitive than our ALMA data at 246.2 GHz. Finally, the maximum recoverable scale (MRS) of the different observations can also play a role. Whilst the MRS of our ALMA observations is $\sim 9''$ (3500 au), the MRS of the VANDAM observations are $\sim 7''$ (2700 au) at 333 GHz and $\sim 5''$ (2100 au) at 32.9 GHz. Our 246.2 GHz data are thus likely filtering less emission compared to those at 333 and 32.9 GHz. In summary, the sources have very different morphologies, therefore constituting a very interesting zoo.

\begin{table*}[ht]
 \caption{The 16 analysed sources, their coordinates, and the known parameters from the \textit{Herschel} HOPS observations.}
    \label{tab:new_sources}
    \centering
    \resizebox{\linewidth}{!}{
    \begin{tabular}{lcccccccl}
    \hline\hline
   \multirow{2}{*}{Source}&R.A.  & Dec. &\multirow{2}{*}{HOPS name \tablefootmark{a,b}} & \multirow{2}{*}{\textit{Herschel} classification\tablefootmark{b}}&$L_{\text{bol}}$\tablefootmark{b} &$T_{\text{bol}}$\tablefootmark{b}&$M_{\text{env}}$\tablefootmark{b}& \multirow{2}{*}{Notes}\\ 
   &[J2000]&[J2000]&&&[\lsol] &[K]&[\msol]& \\
    \hline
    CSO33-a &05:35:19.41 &$-$05:15:38.41&HOPS-56-B  &0&23.3&48.1&0.8&\\
    CSO33-b &05:35:19.48 &$-$05:15:33.08&HOPS-56-A-A/B/C  &0&23.3&48.1&0.8&triple system \tablefootmark{}\\
    CSO33-c & 05:35:19.81&$-$05:15:35.22&V2358 Ori  &0&23.3&48.1&0.8&\\
    FIR6c-a &05:35:21.36&$-$05:13:17.85  &HOPS-409 &0&8.2&28.4&1.8&\\
    FIR2& 05:35:24.30& $-$05:08:30.74 &HOPS-68  &I&5.7&100.6&1.0&\\
    FIR1a-a &05:35:24.87 &$-$05:07:54.63 &HOPS-394-B  & 0&6.6&45.5&0.2&\\
    FIR1a-b &05:35:24.05 &$-$05:07:52.07 &HOPS-394-A  &0&6.6&45.5&0.2&\\
     MMS9-a &05:35:25.97 &$-$05:05:43.34 &HOPS-78-A  &0&8.9&38.1&0.2&\\
    MMS9-b &05:35:26.15 &$-$05:05:45.80 &HOPS-78-B & 0&8.9&38.1&0.2&\\
    MMS9-c &05:35:26.18&$-$05:05:47.14  & HOPS-78-C &0&8.9&38.1&0.2&\\
    MMS9-d & 05:35:25.92&$-$05:05:47.70 & HOPS-78-D &0&8.9&38.1&0.2&no VLA counterpart\\
    MMS5 &05:35:22.47 &$-$05:01:14.34 & HOPS-88&0 &15.8&42.4&0.4&\\
     MMS2-a &05:35:18.34&$-$05:00:32.96  & HOPS-92-A-A/B & Flat\tablefootmark{d}&20.1&186.3&5.6 & binary\tablefootmark{c}\\
    MMS2-b & 05:35:18.27& $-$05:00:33.95 &HOPS-92-B & Flat&20.1&186.3&5.6&\\
    CSO3-b &05:35:16.17 &$-$05:00:02.50  &HOPS-94 &I&6.7&123&0.1& no 333 GHz  counterpart\\
    SIMBA-a &05:35:29.72 &$-$04:58:48.60&HOPS-96  &0&6.2&35.6&1.9&\\
    \hline
    \end{tabular}}
    \tablefoot{
    \tablefoottext{a}{\citealt{fischer2013}}
\tablefoottext{b}{\citealt{furlan2016}}
\tablefoottext{c}{\citealt{tobin2020}}
\tablefoottext{d}{Protostars characterised by a flat spectral energy distribution (SED) in $\lambda F_\lambda$ from $\sim$2 $\mu$m to 24 $\mu$m.}

}

\end{table*}

\subsection{Flux densities and source size derivation}\label{subsec:sed}

In order to derive a meaningful dust SED, the flux density at each frequency (333, 246.2, and 32.9 GHz) has to be measured with the same source size. 
Considering that we have different sets of observations with different angular resolutions and in order to remain consistent in the size choice when measuring the flux density at each wavelength and for each source, we used the method described below.

We used the ORANGES data set to perform a fit of the  16 sources in the visibility plane with the task \textit{uv\_fit} in MAPPING from the GILDAS package.
We chose to fit each source with either a point source or a simple 2D Gaussian function, circular or elliptical depending on what was the best fit. The source size (i.e. the size of the fitted component) is given as an output of the fit, except when using the point source function. All visibilities are considered, but a single-component fit corresponds to the most compact and intense emission component of the sources. The nature of the fitted component is likely the disk + inner envelope (warmest and densest part of the envelope). The extended envelope, which makes only a small contribution, remains thus in the fit residuals. For the sources fitted with a point source function we used the deconvolved size given at 333 GHz (or 32.9 GHz if not available at 333 GHz) in \citet{tobin2020} since these observations have a smaller beam than ours. Once the size for each source is determined, we convolved it with the beam size of each data set (333 GHz, 246.2 GHz and 32.9 GHz; see Table \ref{tab:beam_param} using the formula $\theta_{obs}=\sqrt{\theta_{beam}^2 + \theta_{source}^2}$;   this resulted in an ellipse of size $\theta_{obs-MAJ} \times \theta_{obs-MIN} $ that we used on the image plane to measure the flux density with the procedure \textit{flux} in MAPPING. An example of uv fit result for one of the source samples is shown in Fig. \ref{fig:c2_res}. We verified a posteriori that the flux density given by the uv fit and the one measured on the image plane are consistent within ~30\% at 246.2 GHz. This is due to the difference between the area under the full width at half maximum (FWHM) of a Gaussian and the area under a total Gaussian, in which the flux densities are measured on the image plane and in the visibility plane respectively. Some of our sources show a difference of up to 50\%, likely because those sources have the most extended continuum emission.

The flux density uncertainties include a 10\% amplitude calibration error. For the derived source sizes, we used the 10\% calibration error even if the size was taken from \citet{tobin2020} since they estimated their calibration to be better than 10\% as well. The errors given by the fit in the visibility plane are negligible as they tend to be less than 1\%.  The derived flux densities at 333, 246.2, and 32.9 GHz, the derived source sizes, and the fitting results are summarised in Table \ref{tab:fluxes}.

\subsection{Specific cases}\label{subsec:spe_cases}

Several specific cases arose during the source fitting. First, for CSO33-a, FIR1a-b, MMS9-a, and MMS9-b we could not get a satisfactory fit for the sources. We thus used the deconvolved sizes given in \citet{tobin2020}. Second, the source FIR1a-a could be fitted only with a point source function. However, no deconvolved size was available in \citet{tobin2020}. We then fitted the source with an arbitrary chosen size, covering the whole compact emission at 246.2 GHz (corresponding to about a $ 50\sigma$ contour). Last, CSO33-b is a marginally resolved triple system at 246.2 GHz, whilst it is resolved at 333 GHz. We chose to fit the triple system with an arbitrary size that was large enough to include the three components of the system. Hence, for FIR1a-a and CSO33-b, caution was  taken for the analysis of their results as the sizes chosen were very likely overestimated, which would therefore lead to underestimating the dust optical depths, H$_2$ column densities, and the (envelope + disk) masses. For the same reason, the dust emissivity spectral indexes for those two sources do not necessarily represent the real values. 

\begin{table*}[ht]
    \centering
        \caption{Measured flux densities and source sizes of the 16 sources in Table \ref{tab:new_sources}. Flux density uncertainties are 10\%. }
    \label{tab:fluxes}
    \begin{threeparttable}
    \resizebox{\linewidth}{!}{
    \begin{tabular}{lcccccl}
    \hline \hline
       \multirow{2}{*}{Source}  & $F_\nu$ (333 GHz) & $F_\nu$ (246.2 GHz) &$F_\nu$ (32.9 GHz)& Decon. Size at 246.2 GHz &Decon. P.A&Notes \\
       & [mJy] &[mJy] & [mJy] &  [$''$] &[$^{\circ}$] & \\
       \hline
        CSO33-a & 70  & 32  & 0.23  & 0.79 $\times$ 0.39 \tnote{b} & 61.1&\\
        CSO33-b & 31.0  & 10  &  0.12  & 0.60 $\times$ 0.60 & ...&The flux densities include the triple system \\
        CSO33-c & 1.5 & 0.9 & 0.044 & 0.07 $\times$ 0.07 & ... & \\
        FIR6c-a & 93  & 44  &  0.7 & 0.31 $\times$ 0.13  & $-$64 & \\
        FIR2    & 65 & 28  & 0.41 & 0.17 $\times$ 0.17 & ...& \\
        FIR1a-a & 14 & 7 & 0.09 & 0.23 $\times$ 0.20 & 36.2 & The flux densities correspond to a $\sim 50\sigma$ contour at 1.3mm\\
        FIR1a-b & 11 & 4.7 &  0.20 & 0.25 $\times$ 0.13\tnote{b} & 80.7& \\
        MMS9-a  & 164  & 72 &  0.8 & 0.44 $\times$ 0.14\tnote{b} & 171.1 & \\
        MMS9-b  & 4.6 & 2.4 &  0.05 & 0.05 $\times$ 0.05\tnote{b} & 179.4&\\
        MMS9-c  & 4.6 & 2.0 &  0.031 & 0.10 $\times$ 0.10 & ... & \\
        MMS9-d  & 3.5 & 2.0 &  $\leq$0.02\tnote{a} & 0.13 $\times$ 0.13 & ... & \\
        MMS5    & 90 & 39  & 0.6 & 0.15 $\times$ 0.13 & $-$26.3 &\\
        MMS2-a  & 74 & 33 & 0.24 & 0.13 $\times$ 0.13 & $-$40.0 & The flux densities include the binary system\\
        MMS2-b  & 86 & 43 &  0.5 & 0.17 $\times$ 0.12 & $-$6.7& \\
        CSO3-b  & $\leq$1\tnote{a} & 0.6 &  0.05 & 0.03 $\times$ 0.02\tnote{c} & 49.2 &  \\
        SIMBA-a & 108 & 53 & 0.6 & 0.13 $\times$ 0.11 & $-$47.2&  \\
       \hline
    \end{tabular}
    }
    \tablefoot{
\tablefoottext{a}{ 3$\sigma$ upper limit.}
\tablefoottext{b}{Deconvolved size at 333 GHz and associated P.A. are from \citet{tobin2020}}
\tablefoottext{c}{Deconvolved size at 32.9 GHz and associated P.A. are from \citet{tobin2020}}
    }
    \end{threeparttable}
\end{table*}

\section{Physical parameters of the sources}\label{sec:properties}

\subsection{Source multiplicity}

\begin{figure}[ht]
    \centering
    \includegraphics[width=\linewidth]{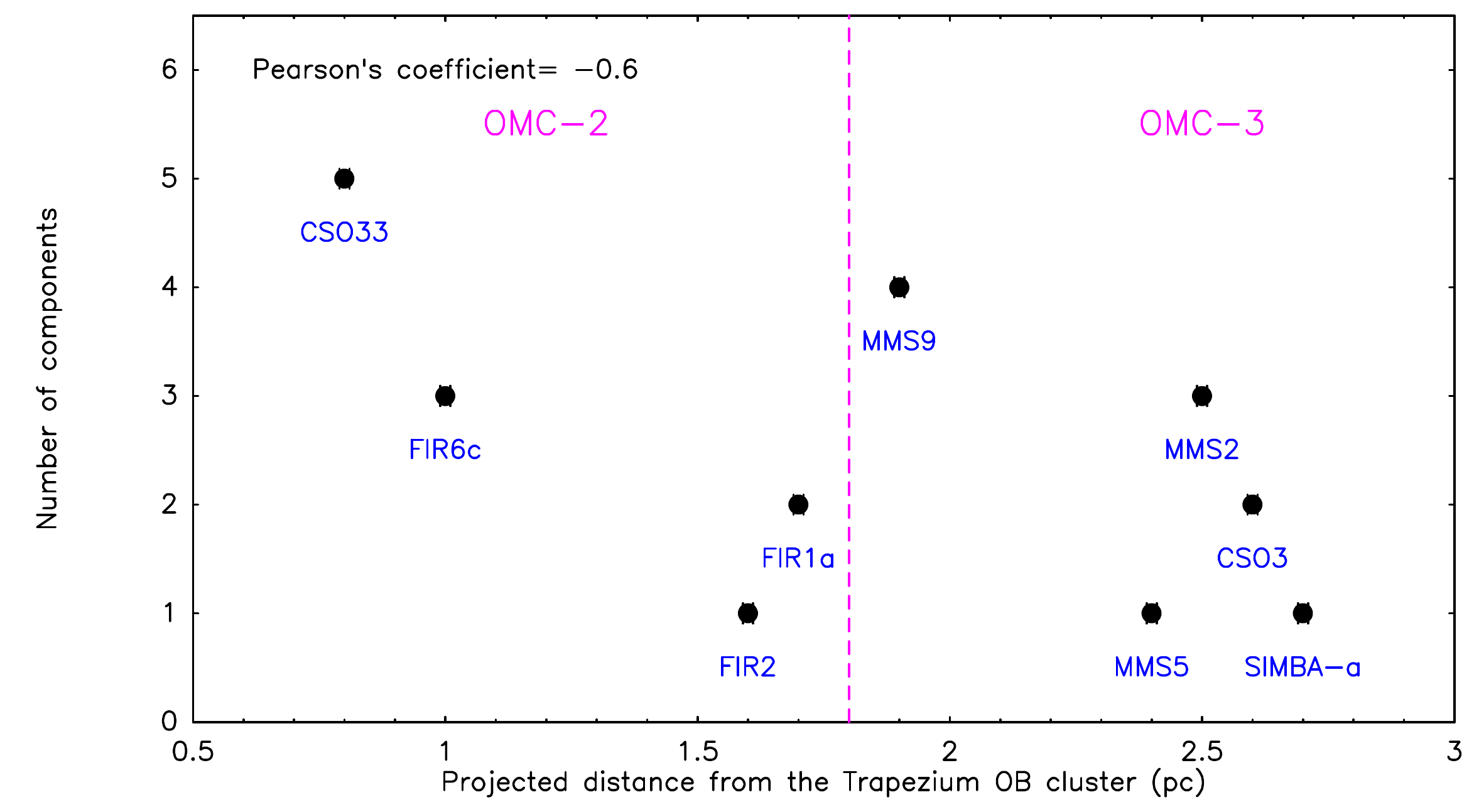}
    \caption{Number of protostars in each of the nine targeted fields (Table \ref{tab:beam_param}) as a function of the projected distance from the Trapezium  OB cluster in pc. The magenta line separates the sources in the OMC-2 and OMC-3 filaments. }
    \label{fig:mult-dist}
\end{figure}

In the nine initially targeted fields we found six multiple systems at a spatial resolution of 0.25$''$ ($\sim\,$100 au), which is  67\% of the source sample (with a statistical error of 33\%). Among these multiple systems, there are three binaries (CSO3, MMS2, and FIR1a), two triple systems (CSO33 and FIR6c), and one quadruple system (MMS9). Our resolution does not allow us to disentangle the triple system of CSO33-b and the binary system of MMS2-a (linear separation less than 100 au), which are both resolved in the VANDAM survey with 0.1$''$ resolution \citep{tobin2020}. 

Other studies have been performed in order to assess the multiplicity of  Class 0 and I sources. On average, a multiplicity percentage of 64\% is found among Class 0 protostars with linear separations in the range 50 - 5000 au (e.g. \citealt{looney2000, maury2010, enoch2011, tobin2013}), whilst it is between 18\% and 47\% for Class I sources with linear separations in the range 45 - 5000 au \citep{haisch2004, duchene2004, duchene2007, connelley2008a, connelley2008b}. 

Therefore, if we take into account only systems with linear separations in the range 45 $-$ 5000 au, we find a multiplicity of 67\% (with a statistical error of 40\%) for Class 0 sources, consistent with that of the other studies. However, for Class I sources, since we do not consider the source CSO3 because the projected separation between the two components is larger than 5000 au, our sample of two sources is statistically too poor to be compared with the other studies.  

 Finally, we found a marginal anti-correlation between the number of protostars in each system and the distance of the system in the OMC-2/3 filament from the Trapezium OB cluster, with a Pearson correlation coefficient of -0.58 (or 34\% probability) as shown in Fig. \ref{fig:mult-dist}. The multiplicity of the sources seems thus to be higher in the southern part of the filament than in the northern part. However, our source sample is rather small, which leads to high statistical uncertainties. Obviously, a study with a larger sample of sources is needed to confirm or disprove these two results. 

\subsection{Method to derive the dust properties and free-free emission }\label{subsec:method}

From the measured flux densities and source size in \S ~\ref{subsec:sed}, we derived the dust emissivity spectral index ($\beta$) and the free-free contribution, the dust optical depth, the H$_2$ column density, and the (envelope + disk) mass using the SED method as follows.

The flux density per beam (i.e. each dust SED point) is given by the  equation
\begin{equation}\label{eq:flux}
    F_\nu=B_\nu(T_\text{d})\left(1-e^{-\tau_\nu}\right)\Omega_s
,\end{equation}
where  $F_\nu$ is  the flux density, $\Omega_s$ is the source solid angle, and $B_\nu(T_\text{d})$ is the Planck function which depends on the dust temperature, $T_{\text{d}}$, as follows:
\begin{equation}\label{eq: B}
B_\nu(T_\text{d})= \frac{2h\nu^3}{c^2}\frac{1}{e^{\frac{hv}{k_{\text{B}}T_\text{d}}}-1}
.\end{equation}
 The dust optical depth $\tau_\nu$ is defined as
\begin{equation}\label{eq:tau}
    \tau_{\nu}=\mu ~m_{\text{H}} ~N_{\text{H2}} ~\text{R} ~\kappa_0\left(\frac{\nu}{\nu_0}\right)^{\beta}
,\end{equation}
where $\mu$=2.8 is the mean molecular weight of the molecular cloud per hydrogen molecule \citep{kauffmann2008}, $m_H$ is the mass of the hydrogen atom, R=0.01 is the mass dust-to-gas ratio, and $\kappa_0=0.9$ cm$^2$g$^{-1}$ is the emissivity of the dust grain at the reference frequency of $\nu_0=230$ GHz \citep{OH94}.

In the following we  indicate as $T_{\text{SED}}$ the dust temperature derived from the SED method, according to Eqs. \ref{eq:flux} to \ref{eq:tau}, to distinguish it from other ways to derive $T_{d}$, described below at {Step 6} (see also Table \ref{tab:temps}).

\paragraph{Step 1, derivation of $\beta$ and $T_\text{SED}$:}
From Eq. (\ref{eq:flux}) we derived the dust emissivity spectral index between 333 and 246.2 GHz ($\beta_{333-246.2}$), 333 and 32.9 GHz ($\beta_{333-32.9}$), and 246.2 and 32.9 GHz ($\beta_{246.2-32.9}$) using Eq. \ref{eq:beta}. As we do not know the sources dust temperature, we derived the indexes for $T_{\text{SED}}$ ranging from 10 to 200 K, with steps of 10 K between 10 and 100 K and steps of 20 K between 100 and 200 K:

\begin{equation}\label{eq:beta}
    {\left(\frac{\nu_1}{\nu_2}\right)}^\beta=\frac{\text{ln}\left(1-\frac{F_{\nu1}}{B_{\nu1}(T_\text{SED})\Omega_s}\right)}{\text{ln}\left(1-\frac{F_{\nu2}}{B_{\nu2}(T_\text{SED})\Omega_s}\right)}.
\end{equation}

For sources without clear free-free emission (see Step 2), we calculated the weighted mean between $\beta_{333-32.9}$ and $\beta_{246.2-32.9}$ for each $T_{\text{d}}$ to derive the final used $\beta$. We note that we did not use $\beta_{333-246.2}$ because of its fairly large uncertainty. For sources with free-free emission, we adopted $\beta$ equal to $\beta_{333-246.2}$. In some cases we were able to constrain the dust temperature to be larger than a minimum temperature because the observed flux density would become lower than the predicted value,  $F_\nu \leq B_\nu(T_\text{d}) ~\Omega_s$, at lower temperatures. 

\paragraph{Step 2, presence of free-free emission:}
From the dust emissivity spectral index derived in Step 1 we can assess whether there is free-free emission. 
The derived $\beta_{333-32.9}$, $\beta_{333-32.9}$, and $\beta_{246.2-32.9}$ only differ when free-free emission at 32.9 GHz is present (where we assume that free-free emission is not present at 246.2 and 333 GHz). 
In that case, $\beta_{333-246.2}$ is expected to be larger than $\beta_{333-32.9}$ and $\beta_{246.2-32.9}$ since a free-free contribution leads to smaller $\beta$ values. 
Therefore, sources presenting a variation in $\beta$ in the different frequencies are those with free-free emission. 

\paragraph{Step 3, derivation of $\tau$:}
We derived the dust optical depth, $\tau_\nu$, as a function of $(T_\text{SED})$ by inverting Eq. (\ref{eq:flux}):
\begin{equation}
    \tau_\nu=-\text{ln}\left(1-\frac{F_\nu}{B_\nu(T_\text{SED})\Omega_s}\right).
\end{equation}
For sources with no free-free emission at 32.9 GHz, we calculated the dust optical depth at each frequency. For sources with free-free contribution at 32.9 GHz we did not calculate the optical depth at this frequency. 

\paragraph{Step 4, derivation of $N_{\text{H2}}$ and $M_{\text{env+disk}}$:}
We derived the H$_2$ column density, $N_{\text{H2}}$, as a function of the SED temperature using Eq. (\ref{eq:tau}). 
For sources without free-free contribution we calculated $N_{\text{H2}}$ at each frequency;  we used the weighted mean of $\beta$ derived in Step 1, which is a function of $(T_\text{SED})$. 
We then averaged the derived $N_{\text{H2}}$ values between the three frequencies to get a final $N_{\text{H2}}$ as a function of the SED temperature. 
For sources with centimetre excess we did not calculate $N_{\text{H2}}$ at 32.9 GHz, and we used $\beta_{333-246.2}$. 
We then averaged the $N_{\text{H2}}$ values between 333 and 246.2 GHz to get a final $N_{\text{H2}}$ as a function of the SED temperature. 

Finally, we derived the (envelope + disk) mass of the sources, $M_{\text{env+disk}}$. We note that following the derivation of the source size,  {(envelope + disk) mass} refers to the mass of the disk and of the immediate and warm envelope (densest and most compact component of the envelope). We used the equation
\begin{equation}
    M_{\text{env+disk}}=\mu m_{\text{H}}N_{\text{H2}}d^2\Omega ,
\end{equation}
where \textit{d} is the distance of the OMC-2/3 filament.

\paragraph{Step 5, derivation of the fraction of free-free emission at 32.9 GHz, FF:}
 Having derived the dust emissivity spectral index (Step 1) and  $N_{\text{H2}}$ (Step 4) and using Eqs. \ref{eq:flux} and \ref{eq:tau}, we can quantify the percentage of free-free emission FF as a function of the SED temperature for each source. 
 To that end, we calculated the flux density at 32.9 GHz that we should have measured if the emission were only due to thermal dust emission, and then compared it to the flux density we actually measured. 
 We therefore derived the percentage of free-free emission at 32.9 GHz as a function of the SED temperature for each source. 

\paragraph{Step 6, derivation of the dust radiative temperature $T_{\text{rad}}$:}
For the sources that are not multiple systems and for which the bolometric luminosity $L_{\text{bol}}$ can be trusted, we computed the dust radiative temperature, $T_{\text{rad}}$, defined as the temperature of a dust grain radiatively heated by a $L_{\text{bol}}$ source and at a distance r from the heating source. 
The following formula, roughly valid for an optically thick dust spherical envelope, provides a good approximate value of $T_{\text{rad}}$ \citep[][Eq. 1]{ceccarelli2000a}:
\begin{equation}\label{eq:Trad}
    T_{\text{rad}}=101\ \text{K}\left(\frac{L_{\text{bol}}}{10\text{\lsol}}\right)^{1/4}\left(\frac{r}{50 \ \text{au}}\right)^{-1/2}
.\end{equation}
In our case, r is the radius of the dust continuum emission source (Table \ref{tab:fluxes}).
In the case of elliptic sources, we did the calculations for both the major and minor axis to get a range of possible values for $T_{rad}$. 

The $T_{\text{rad}}$ computed in this way provides an approximate value of the dust temperature at the border of the continuum emission source and, if the emission is optically thick, it would be approximately the measured dust temperature.
Therefore, it can be used to have an approximate value of the dust temperature in the cases where it could not be otherwise constrained (see Step 1).

\paragraph{Step 7, construction of dust SEDs from the derived dust parameters. }
Once we have constrained the dust temperature, the dust emissivity spectral index, and the H$_2$ column density ranges for each of our source samples, we can plot the dust SEDs corresponding to those derived parameters. To this end, we used   Eqs. \ref{eq:flux}, \ref{eq: B}, and \ref{eq:tau} to calculate the flux density associated with these parameters as a function of frequency, within the dust temperature range derived from the Step 1. When they were available we also used $T_{\text{rad}}$ and $T_{\text{bol}}$.  We used frequencies in the range [10, 340] GHz, with steps of 10 GHz bteween 10 and 100 GHz and with steps of 20 GHz between 100 and 340 GHz. This results in a range of possible dust SEDs for a given set of dust parameters ($T_{d}$, $\beta$, and $N_{\text{H2}}$). To compare the calculated flux densities with those we measured, and as a self-consistency check, we added the latter on the dust SED plots. 

\paragraph{Summary:}
We used the SED method to derive $\beta$ (Step 1), $\tau_\nu$ (Step 3), $N_{\text{H2}}$ and $M_{\text{env+disk}}$ (Step 4), and FF (Step 2 and 5). 
In the analysis, we used two additional temperatures (see Table \ref{tab:temps}): the bolometric temperature, $T_{\text{bol}}$, derived by \citet{furlan2016}, and the dust radiative temperature, $T_{\text{rad}}$, derived from the bolometric luminosity of each source and its size (Step 6). Finally, we used these derived parameters to build the associated SEDs (Step 7).
All uncertainties in the derivation of $\beta, \tau_\nu$, $N_{\text{H2}}$, and $M_{\text{env+disk}}$ are detailed in Appendix \ref{appdx:uncertainties}. In the following  subsections we report the results from this SED fitting technique.

\subsection{Dust temperature}\label{subsec:results_temps}

\begin{table*}[ht!]
    \caption{Definition of each of the temperatures used in this study}
    \label{tab:temps}
    \centering
    \resizebox{0.95\linewidth}{!}{%
    \begin{tabular}{lcl}
    \hline \hline
       Temperature  & Symbol &  Definition \\
       \hline
         SED temperature & $T_{\text{SED}}$ & Dust temperature constrained from the SED method over the frequency range 32.9-333 GHz.\\
         Bolometric temperature & $T_{\text{bol}}$ & Temperature of a blackbody with the same mean frequency as the observed SED.\\
         Dust radiative temperature & $T_{\text{rad}}$&  Dust temperature of  a dust grain radiatively heated by a $L_{\text{bol}}$ source and at a distance r from the heating source.\\
         \hline
    \end{tabular}
    }
\end{table*}

\paragraph{$T_{\text{SED}}$:}
In order to constrain the dust temperature range we took into account that $\beta$ should range between 0 and 2, as discussed in \S ~\ref{sec:review}.
We thus constrained the range of dust temperatures for which $\beta$ respects this condition (see Fig. \ref{fig:beta_T_1}). The derived temperature range, $T_{\text{SED}}$, in each source is reported in Table \ref{tab:fit_results}.

\paragraph{$T_{\text{bol}}$ and $T_{\text{rad}}$:}
Five sources of our sample possess an estimate of the bolometric luminosity and are single systems within the spatial resolution of 0.$''$1 ($\sim 40$ au): FIR2, MMS5, SIMBA-a, CSO3-b, and FIR6c-a.
Although three sources lie in the FIR6c-a Herschel beam, from which the bolometric luminosity is derived, we note that the other two are supposedly pre-stellar cores \citep{kainunlainen2017}. Therefore, the bolometric luminosity is reliably attributable only to FIR6c-a. 

For these five sources we derived the bolometric and radiative temperatures, reported in Table \ref{tab:other_temp}, and the other dust properties at these temperatures.
Figures \ref{fig:beta_T_1} and \ref{fig:FF_T_1}  to \ref{fig:mass_T_1} show how these temperatures compare with the dust temperature derived from the SED. In FIR2 and CSO3-b, $T_{\text{bol}}$ lies within the range of dust temperatures that we derived from the SED. 
However, in the other three sources, FIR6c-a, MMS5, and SIMBA-a, $T_{\text{bol}}$ is lower.
This can be explained by the fact that these objects have high dust millimetre optical depths $\tau$. 
At the mid-infrared (MIR) wavelengths of \textit{Herschel} and \textit{Spitzer} the dust becomes optically thick at a larger radius than that at millimetre wavelengths; consequently, the dust probed by the MIR is colder than that probed by  millimetre wavelengths. 

On the contrary, with the exception of SIMBA-a and CSO3-b, the radiative temperatures are consistent with the range derived from the SED. They are reported in Table \ref{tab:other_temp}, as are the corresponding dust properties and free-free excess. For CSO3-b we used the bolometric temperature and the associated dust parameters for the analysis.
Finally, SIMBA-a possess a radiative temperature that is lower than that derived from the SED; in this case we could not further constrain the temperature and we used the results from the SED. 

\paragraph{Hot, warm, and cold sources:}
We classified the sources in four categories: (1) cold ($T_{\text{d}} < 50 $ K), (2) warm ($T_{\text{d}} \geq 50 $ K), (3) hot ($T_{\text{d}} \geq 90 $ K), and (4) unknown (sources for which the range spans from $\leq 50 $ K to $\geq 90 $ K. 
Following this classification,  FIR6c-a, FIR2, MMS5, MMS2-a, CSO3-b, and SIMBA-a are hot sources, while FIR1a-a and MMS9-d are cold sources. In between, MMS9-a,b and MMS2-b are classified as warm sources. 
Finally, for CSO33-a,-b,-c, FIR1a-b, and MMS9-c, the range of dust temperatures could not be constrained.

\subsection{Dust emissivity spectral index and free-free presence}\label{subsec:results-beta}

\begin{figure*}[ht]
    \centering
    \includegraphics[width=0.9\linewidth]{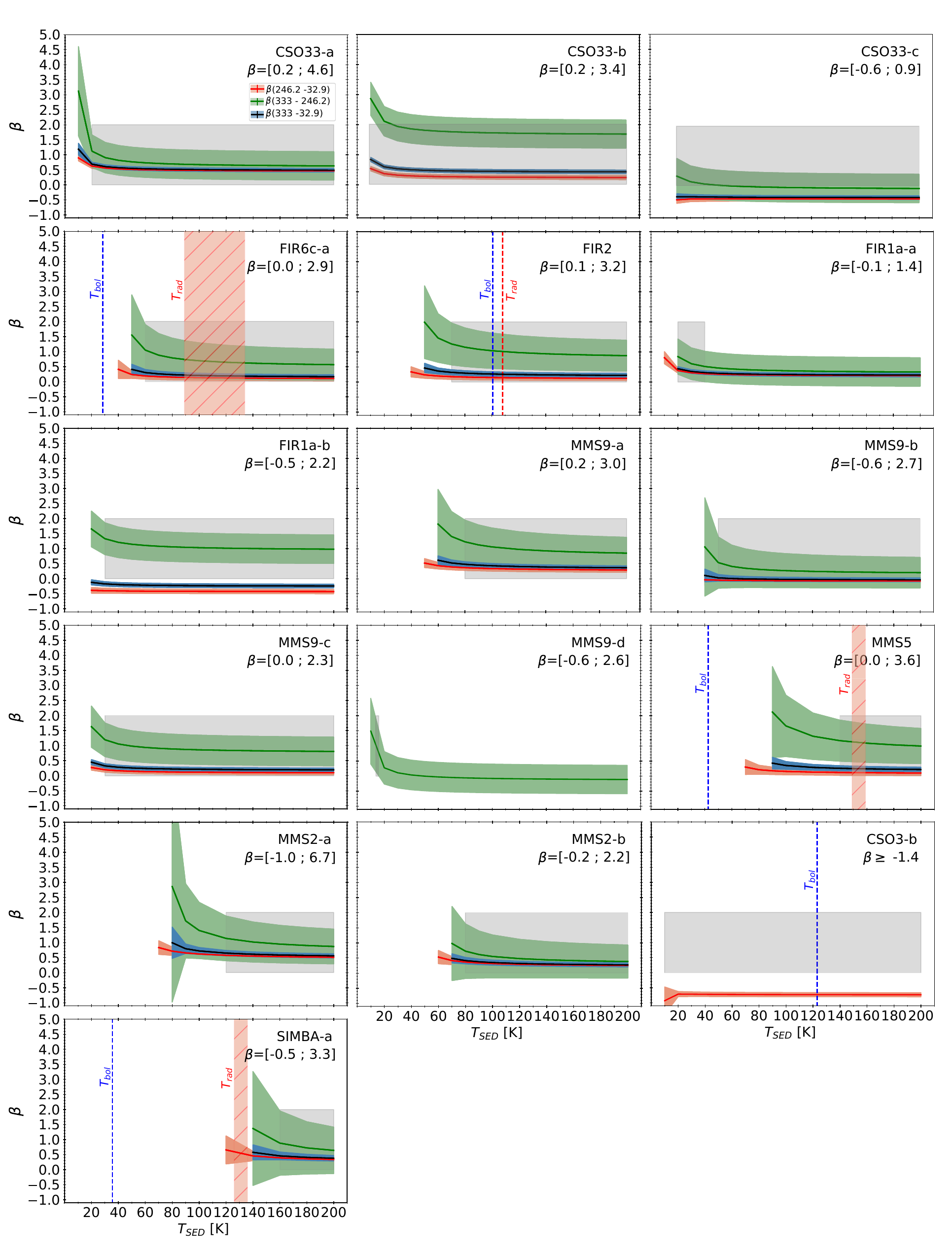}
    \caption{Plots of dust emissivity spectral indexes, $\beta$, as a function of the SED temperature for each source. The indexes were   derived between 333 GHz and 32.9 GHz (black lines), 333 GHz and 246.2 GHz (green lines), and 246.2 GHz and 32.9 GHz (red lines). The shaded grey area indicates the authorised range of $\beta$ values (0--2) provided by theoretical works (see \S ~\ref{sec:review}). Applying these limits on $\beta$ constrains the possible range of $T_{SED}$ values. For each source, the minimum and maximum values of $\beta$ derived over the full temperature range are indicated at the top right of the plots. For single systems, the bolometric temperature, $T_{bol}$, derived from \textit{Herschel} \citep{furlan2016}, and the radiative temperature, $T_{rad}$, are indicated by a vertical blue dashed line and red shaded area, respectively.}
    \label{fig:beta_T_1}
\end{figure*}

Figure \ref{fig:beta_T_1} shows the derived $\beta$ as a function of the SED temperature for each of the sources. Table \ref{tab:fit_results} lists the dust emissivity spectral indexes derived between 333 and 246.2 GHz, 333 and 32.9 GHz, and 246.2 and 32.9 GHz within the constrained range of dust temperatures. 
As discussed in the previous subsection, we only consider the indexes lying in the range [0;2] in the following analysis. 
The final range of $\beta$ values of each source used in the following are reported in Tables \ref{tab:fit_results_main} and \ref{tab:fit_results}.
The majority of the sources (9/16) have a dust emissivity spectral index $< 1$.

In five sources (CSO33-b, FIR2, FIR1a-b, MMS9-c, and MMS5), the index between 333 and 246.2 GHz (green) is larger than the indexes between 246.2 and 32.9 GHz (red) and 333 and 32.9 GHz (black) across the derived SED temperature range. Those sources thus present   free-free emission. 
For FIR6c-a and MMS9-a, uncertainties on $\beta_{333-246.2}$ are too large to definitively assess the presence of free-free emission at 32.9 GHz.

For the remaining sources, the three derived $\beta$ are consistent with each other within the error bars and are thus considered as sources without free-free emission. The results are reported in Table \ref{tab:fit_results_main}. 

\subsection{Opacities and envelope plus disk masses}

The derived dust optical depths vary between 0.01 and 2.4 at 333 GHz, between 0.004 and 1.6 at 246.2 GHz, and between 0.01 and 1.3 at 32.9 GHz respectively. 
The derived H$_2$ column densities range between $1 \times 10^{23}$ \pcm \ and $3.8 \times 10^{25}$ \pcm.
Finally, the derived (envelope + disk) masses range between $2\times10^{-4}$\msol \ and 0.1\msol. 
All the derived values are reported in Table \ref{tab:fit_results_main}, while the plots of these parameters as a function of the dust temperature are shown in Figs. \ref{fig:tau_T_1}, \ref{fig:nh2_T_1}, and \ref{fig:mass_T_1}.

\subsection{Dust SEDs}

\begin{figure*}
    \centering
    \includegraphics[width=\linewidth]{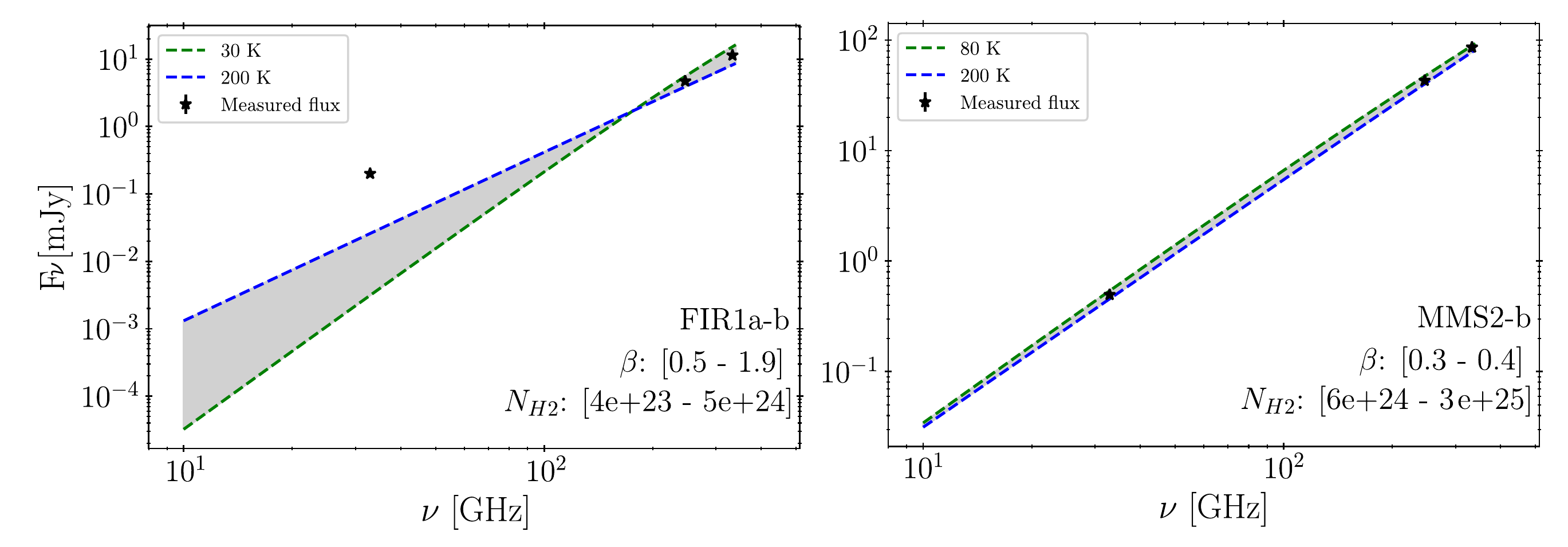}
    \caption{Example of dust SEDs obtained for two sources, FIR1a-b (left) and MMS2-b (right). The dust SEDs, calculated for the corresponding dust temperature range of each source and from the final parameters derived for each source ($\beta$, $N_{\text{H2}}$, are shown by the grey shaded area. These parameters are indicated on the lower right side of the plot. Measured flux densities are represented by black filled stars.  The dust SEDs corresponding to the two extreme values of the dust temperature range are represented by the blue and green dashed lines.}
    \label{fig:ex_SEDs-plots}
\end{figure*}

The final dust SEDs corresponding to the derived dust parameters for each source are shown in Fig. \ref{fig:SEDS_plot}. We show in Fig. \ref{fig:ex_SEDs-plots} an example of two dust SED plots, for FIR1a-b and MMS2-b. In FIR1a-b we see that the flux density at 32.9 GHz is above the dust SEDs lines, indicating  the presence of free-free emission, while this is not the case in MMS2-b.

\subsection{Fraction of free-free emission at 32.9 GHz} 
As said in \S ~\ref{subsec:results-beta}, a $\beta_{333-246.2}$ larger than the other two derived values of beta implies the presence of free-free emission at 32.9 GHz.
In the five sources where this
happens (CSO33-b, FIR2, FIR1a-b, MMS9-c, and MMS5) we calculated the percentage fraction of the 32.9 GHz emission from free-free with respect to the dust continuum emission. 
The derived values are reported in Table \ref{tab:fit_results_main} and shown in Fig. \ref{fig:FF_T_1}. 
We can see that for these sources the derived free-free emission ranges between 47\% and 100\%, whilst in sources with a priori no free-free emission, the largest percentage that can be attributed to free-free emission is 24\%. 
We note that no 333 GHz emission is detected towards CSO3-b and, consequently, we cannot assess whether an excess at 32.9 GHz is present.

\begin{table*}[ht]
    \centering
     \caption{Results for the dust parameters derived for each source. First, the linear source sizes are reported in the  Col. 2. Then, the dust temperature range and the corresponding category are reported in Cols. 3 and 4 respectively. The dust emissivity spectral indexes are reported in Col. 5. The range of dust optical depth values derived at 333 GHz, $\tau_{333}$, 246.2 GHz, $\tau_{246.2}$, and 32.9 GHz, $\tau_{32.9}$, as well as the H$_2$ column densities, $N_{\text{H2-mm}}$, and (envelope + disk) masses, $M_{\text{env+disk}}$, corresponding to the derived range of temperature, are reported in Cols. 6 to 8. The range of free-free emission percentage at 32.9 GHz is indicated in Col. 9. In the last column we report the new classification of the sources.}  \label{tab:fit_results_main}
    \resizebox{0.95\linewidth}{!}{%
    \begin{tabular}{lccccccccc}
    \hline \hline
    \multirow{4}{*}{Source}& \multirow{3}{*}{Size}&\multirow{3}{*}{$T_{\text{d}}$}&\multirow{4}{*}{Category}&\multirow{4}{*}{$\beta$}&$\tau_{333}$&\multirow{3}{*}{$N_{\text{H2-mm}}$}&\multirow{3}{*}{$M_{\text{env+disk}}$}& \multirow{3}{*}{FF}&\multirow{4}{*}{New classification}\\
      &&&&&$\tau_{246.2}$&&&&\\
      &&&&&$\tau_{32.9}$&&&&\\
      &[au $\times$ au]&[K]&&&&[$\times 10^{24}$ cm$^{2}$]&[$\times 10^{-2}$\msol]&[\%]&\\
      \hline
         \multirow{3}{*}{CSO33-a} &\multirow{3}{*}{310 $\times$ 153}&\multirow{3}{*}{20 $-$ 200}&\multirow{3}{*}{unknown}&\multirow{3}{*}{0.5 $-$ 0.7}&0.02 $-$ 0.4&\multirow{3}{*}{0.4 $-$ 6}&\multirow{3}{*}{1 $-$ 13}&\multirow{3}{*}{0 - 24}& \multirow{3}{*}{Class 0 or I}  \\
        &&&&&0.01 $-$ 0.3& &&& \\
        &&&&&0.01 $-$ 0.1&&&&\\
        \hline
        \multirow{3}{*}{CSO33-b}&\multirow{3}{*}{236 $\times$ 236}&\multirow{3}{*}{10 $-$ 200} &\multirow{3}{*}{unknown}&\multirow{3}{*}{1.2 $-$ 2.0}&$\geq$ 0.01&\multirow{3}{*}{$\geq$ 0.08}&\multirow{3}{*}{$\geq$ 0.2}&\multirow{3}{*}{88 - 100}&\multirow{3}{*}{Class 0}\\
        &&&&&$\geq$ 0.004&&&& \\
        &&&&&...&&&\\
         \hline
        \multirow{3}{*}{CSO33-c} &\multirow{3}{*}{27.5 $\times$ 27.5}&\multirow{3}{*}{20 $-$ 200}&\multirow{3}{*}{unknown}&\multirow{3}{*}{0}&0.02 $-$ 0.6&\multirow{3}{*}{0.7 $-$ 13} &\multirow{3}{*}{0.02 $-$ 0.5}&\multirow{3}{*}{0 - 21}&\multirow{3}{*}{Class II}\\
        &&&&&0.02 $-$ 0.5&& &&  \\
        &&&&&0.1 $-$ 1.5&&& \\
         \hline
        \multirow{3}{*}{FIR6c-a} &\multirow{3}{*}{122 $\times$ 51}&\multirow{3}{*}{89 $-$ 134}&\multirow{3}{*}{hot}&\multirow{3}{*}{0.1 $-$ 0.3}&0.3 $-$ 0.7 & \multirow{3}{*}{7 $-$ 15} &\multirow{3}{*}{1.5 $-$ 4}&\multirow{3}{*}{0 - 21}&\multirow{3}{*}{Class 0}\\
        &&&&&0.2 $-$ 0.6&&& & \\
        &&&&&0.2 $-$ 0.4&&&&\\
         \hline
        \multirow{3}{*}{FIR2} &\multirow{3}{*}{67 $\times$ 67}&\multirow{3}{*}{101 $-$ 108}&\multirow{3}{*}{hot}&\multirow{3}{*}{0.4 $-$ 1.6} &0.4 $-$ 0.6&\multirow{3}{*}{7 $-$ 10}&\multirow{3}{*}{1 $-$ 2}& \multirow{3}{*}{55 - 100}& \multirow{3}{*}{Class I} \\
        &&&&&0.3 $-$ 0.4&&&& \\
        &&&&&...&&&\\
         \hline
        \multirow{3}{*}{FIR1a-a} &\multirow{3}{*}{90 $\times$ 79}&\multirow{3}{*}{20 $-$ 40}&\multirow{3}{*}{cold}&\multirow{3}{*}{0.3 $-$ 0.4}&$\geq$ 0.1& \multirow{3}{*}{$\geq$ 3 } &\multirow{3}{*}{$\geq$ 1}& \multirow{3}{*}{0 - 22}&\multirow{3}{*}{Class 0 or I} \\
        &&&&&$\geq$ 0.1&&&&\\
        &&&&&$\geq 0.1$&&& \\
         \hline
        \multirow{3}{*}{FIR1a-b} &\multirow{3}{*}{98 $\times$ 51}&\multirow{3}{*}{30 $-$ 200}&\multirow{3}{*}{unknown}&\multirow{3}{*}{0.5 $-$ 1.9}&0.03 $-$ 0.3& \multirow{3}{*}{0.4 $-$ 5 } &\multirow{3}{*}{0.1 $-$ 1}& \multirow{3}{*}{87 - 100}&\multirow{3}{*}{Class 0}  \\
        &&&&&0.02 $-$ 0.2&& && \\
        &&&&&...&&&\\
         \hline
        \multirow{3}{*}{MMS9-a} &\multirow{3}{*}{173 $\times$ 55}&\multirow{3}{*}{80 $-$ 200}&\multirow{3}{*}{warm}&\multirow{3}{*}{0.3 $-$ 0.4}&0.2 $-$ 1.0&\multirow{3}{*}{5 $-$ 19} &\multirow{3}{*}{2 $-$ 7}&\multirow{3}{*}{0 - 23} &\multirow{3}{*}{Class 0}  \\
        &&&&&0.2 $-$ 0.7&&& & \\
        &&&&&0.1 $-$ 0.4&&&\\
         \hline
        \multirow{3}{*}{MMS9-b} &\multirow{3}{*}{20 $\times$ 20}&\multirow{3}{*}{50 $-$ 200}&\multirow{3}{*}{warm}& \multirow{3}{*}{0}&0.1 $-$ 1.5& \multirow{3}{*}{4 $-$ 31}&\multirow{3}{*}{ 0.05 $-$ 0.5}&\multirow{3}{*}{0 - 22}&\multirow{3}{*}{Class 0 or I} \\
        &&&&&0.1 $-$ 1.3&&&& \\
        &&&&&0.2 $-$ 1.3&&&\\
         \hline
        \multirow{3}{*}{MMS9-c} &\multirow{3}{*}{39 $\times$ 39}&\multirow{3}{*}{30 $-$ 200}&\multirow{3}{*}{unknown}& \multirow{3}{*}{0.3 $-$ 1.8}&0.03 $-$ 0.5& \multirow{3}{*}{0.6 $-$ 7.5 } &\multirow{3}{*}{0.03 $-$ 0.5}&\multirow{3}{*}{47 - 100}&\multirow{3}{*}{Class 0 or I}  \\
        &&&&&0.03 $-$ 0.3&&&&\\
        &&&&&... &&&\\
         \hline
        \multirow{3}{*}{MMS9-d} &\multirow{3}{*}{51 $\times$ 51}&\multirow{3}{*}{10 $-$ 20}&\multirow{3}{*}{cold}& \multirow{3}{*}{0 $-$ 2.0}&0.3 $-$ 2.4 & \multirow{3}{*}{5 $-$ 38}&\multirow{3}{*}{0.5 $-$ 4} &\multirow{3}{*}{...}&\multirow{3}{*}{Class II or background object}  \\
        &&&&&0.3 $-$ 1.4 &&& &\\
        &&&&&...&&&  \\
         \hline
        \multirow{3}{*}{MMS5} &\multirow{3}{*}{59 $\times$ 51}&\multirow{3}{*}{149 $-$ 159}&\multirow{3}{*}{hot}&\multirow{3}{*}{0.5 $-$ 1.8}&0.5 $-$ 0.9& \multirow{3}{*}{10 $-$ 15} &\multirow{3}{*}{1 $-$ 2}& \multirow{3}{*}{60 - 100}&\multirow{3}{*}{Class 0 } \\
        &&&&&0.4 $-$ 0.6&&&&  \\
        &&&&&...&&&\\
         \hline
        \multirow{3}{*}{MMS2-a} &\multirow{3}{*}{51 $\times$ 51}&\multirow{3}{*}{120 $-$ 200}&\multirow{3}{*}{hot}&\multirow{3}{*}{0.5 $-$ 0.6}&0.4 $-$ 1.2& \multirow{3}{*}{8 $-$ 20} &\multirow{3}{*}{1 $-$ 2}& \multirow{3}{*}{0 - 23} &\multirow{3}{*}{Class I} \\
        &&&&&0.3 $-$ 0.9&&&&\\
        &&&&&0.1 $-$ 0.3&&&\\
         \hline
        \multirow{3}{*}{MMS2-b} &\multirow{3}{*}{51 $\times$ 47}&\multirow{3}{*}{80 $-$ 200}&\multirow{3}{*}{warm}&\multirow{3}{*}{0.3 $-$ 0.4}&0.3 $-$ 1.7& \multirow{3}{*}{6 $-$ 30} &\multirow{3}{*}{1 $-$ 5}&\multirow{3}{*}{0 - 22} &\multirow{3}{*}{Class I}\\
         &&&&&0.2 $-$ 1.3&&&& \\
        &&&&&0.1 $-$ 0.6&&&\\
         \hline
        \multirow{3}{*}{CSO3-b} &\multirow{3}{*}{...}&\multirow{3}{*}{123\tablefootmark{a}}&\multirow{3}{*}{hot}&\multirow{3}{*}{$\geq 0$}&...&\multirow{3}{*}{$\geq$ 0.4}\tnote{a}&\multirow{3}{*}{$\geq$ 0.02}& \multirow{3}{*}{$\geq 0$}&\multirow{3}{*}{Class I}\\
        &&&&&$\geq$ 0.02 &&&& \\
        &&&&&...&& & \\
         \hline
        \multirow{3}{*}{SIMBA-a} &\multirow{3}{*}{51 $\times$ 43}&\multirow{3}{*}{160 $-$ 200}&\multirow{3}{*}{hot}&\multirow{3}{*}{0.4}&0.7 $-$ 2.2& \multirow{3}{*}{18 $-$ 36}  &\multirow{3}{*}{1 $-$ 3}&\multirow{3}{*}{0 - 23} &\multirow{3}{*}{Class 0}\\
        &&&&&0.6 $-$ 1.6&&&& \\
        &&&&&0.4 $-$ 0.7&&&\\
      \hline
    \end{tabular}%
    }
    \tablefoot{
\tablefoottext{a}{$T_{\text{bol}}$}
    }
\end{table*}

\section{Discussion}\label{sec:discussion}
\subsection{Dust properties}

\paragraph{Dust temperature:}
From the above analysis (\S ~\ref{subsec:results_temps}) six sources are classified as hot ($T_{\text{d}}\geq 90$ K), three as warm ($T_{\text{d}}\geq 50$ K), and two as cold ($T_{\text{d}} < 50$ K). 
We could not constrain the temperature range in the remaining five sources. 
Our derived temperatures are higher than those derived by previous OMC-2/3 surveys, which found temperatures lower than 30 K (e.g. \citealt{chini1997, lis1998, JB1999}) or around 50 K towards the FIR6 region \citep{shimajiri2009}. 
The discrepancy can be due to the fact that we do not probe the same region as that of these studies, and that our higher angular resolution allows us to probe a more internal region of the protostar envelope, which is warmer than the outer one.

\paragraph{Dust emissivity spectral index:}
The derived dust emissivity spectral indexes range between 0 and 2;  the majority (9/16) of the sources have  values $< 1$.  Compared to the canonical value expected for protostar envelopes, between 1 and 2 \citep[e.g.][]{natta2007, jorgensen2007, PC2011, chiang2012, sadavoy2013, forbrich2015, chen2016, li2017, bracco2017,galametz2019}, our range of derived values is lower. 
However, several other studies at scales of 100-2000 au, such as ours, found $\beta \leq 1$ (e.g. \citealt{jorgensen2007, kwon2009, chiang2012, miotello2014, li2017, galametz2019}). 
Different explanations supported by theoretical, modelling, and laboratory works are given in the literature. 

The presence of large grains (millimetre- or centimetre-sized)  due to grain growth, a large optical depth,  and/or dust self-scattering can lead to a decrease in the dust emissivity index in the (sub-)millimetre range (e.g. \citealt{MN1993, OH94, draine2006, jones2013, jones2017, wong2016, ysard2019}). 
However, since we took into account the optical thickness of the dust in the derivation of the dust emissivity spectral index, large optical depths should not affect our estimates. 
An example of the first explanation is that inner protostellar and/or protoplanetary disks, if present, can contribute to the decrease in the dust emissivity index, and the measured dust emissivity spectral indexes in those disks are usually $\beta \leq 1$ \citep[e.g.][]{BS91, ricci2012, perez2012, ubach2012,miotello2014,bracco2017,liu2019,nakatani2020}. 

Alternatively, several laboratory works found an intrinsic dependence between the dust emissivity spectral index and the dust temperature in the millimetre range: when the  dust temperature increases, $\beta$ decreases (e.g. \citealt{dupac2003, mennella98, boudet2005}). Finally, another possible explanation for the low value of $\beta$ is the difference in the dust grain composition compared to that of the diffuse ISM (e.g. \citealt{aanestadt75, agladze1996, mennella98, coupeaud2011, jones2013, jones2017, ysard2019}).

\paragraph{Envelope+disk mass:}
Our estimated (envelope+disk) masses are in good agreement with the values derived by the VANDAM survey \citep{tobin2020}, except for FIR1a-a and MMS9-d. 
In FIR1a-a we only could provide lower limits because we could not derive the source size accurately (see Sect. \ref{subsec:spe_cases}), and they are higher than the values  found in \citet{tobin2020} by at least a factor of 5. The mass derived for MMS9-d differs by up to one order of magnitude compared to what is found in \citet{tobin2020}. In both cases, the discrepancy can be explained by the fact that \citet{tobin2020} assumed an optically thin dust, whilst the dust might actually be optically thick, especially in the case of MMS9-d (see Table \ref{tab:fit_results}). Therefore, the dust mass for those two objects would be underestimated in their study. \footnote{For the sake of completeness, we also compared the estimated mass of our sources with those derived by \textit{Herschel} \citep{furlan2016} and found our masses to be between one and two orders of magnitude lower than those of \textit{Herschel}. This is most probably due to the fact that our observations probe smaller regions than that of the HOPS survey.}

Caution should be taken when using the estimated dust masses as they depend on the choice of the dust opacity coefficient in the millimetre wavelengths, which is known within one order of magnitude accuracy. 
This uncertainty implies about one order of magnitude uncertainty on the disk mass as well \citep{MN1993, natta2004, ricci2010a,fanciullo2020}.

\subsection{Free-free emission}
We clearly found evidence of free-free emission at 32.9 GHz towards 5 out of 16 sources (i.e. $\sim$ 31\% of the source sample). 
We can compare this result with the previous study from \citet{reipurth1999},  who surveyed the OMC-2/3 filament with the VLA at 8.3 GHz (3.6 cm). 
They detected 11 sources that were likely either protostars or very young stars. Of these 11 sources, 2  coincide with some of our multiple systems, namely MMS9 and FIR1a. However, they detected free-free emission towards the MMS2 system whilst we did not, and we detected free-free emission towards FIR2 and MMS5 whilst they did not. We also found free-free emission towards CSO33-b, but this source is not included in their mapped area. 

The non-detection of 8.3 GHz emission towards FIR2 and MMS5 can be due to a difference in sensitivity between the VANDAM survey observations (8 $\mu$Jy/beam sensitivity) and those of \citet{reipurth1999}($\sim$ 40 $\mu$Jy/beam sensitivity). 
In addition, the difference in the frequency of the two surveys can also explain the discrepancy in the results.
Thermal radio jets and/or winds, which are likely to be responsible of the free-free emission in Class 0 and I protostars \citep{anglada1995,anglada1998,  rodriguez97}, show a SED with a positive spectral index at centimetre wavelengths. Therefore, if the free-free emission is weak at 32.9 GHz, it will be even weaker at 8.3 GHz, and thus probably not detectable at this wavelength. 

Finally, a possible explanation to why we did not detect a centimetre excess towards either of the two sources composing the MMS2 system is that the 8.3 GHz emission traces  a non-thermal radio emission associated with a shock in the outflow driven by the system \citep{yu1997}. 
Non-thermal emission at centimetre wavelengths have been detected in YSO jets \citep[e.g.][]{rodriguez1989, marti93, garay96, wilner1999, anglada2018} and is usually found in strong radio knots, away from the core, with a negative spectral index at centimetre wavelengths. Therefore, if the emission is seen at 8.3 GHz, it is not necessarily seen at 32.9 GHz. In the MMS2 field we do not see any other clear 32.9 GHz emission apart from that seen towards the centre of the protostars, which strengthens this hypothesis.

\subsection{Evolutionary status}\label{subsec:evol}
Our sample sources have been previously classified into Class 0, Class I and Flat objects  based on SEDs from \textit{Herschel}, \textit{Spitzer,} and submillimetre APEX photometric observations \citep[see][]{furlan2016}. However, the angular resolution of these observations does not allow us to disentangle multiple systems where components have close projected separations, as is the case for most of our sources. Hence, cross-correlating information between the different parameters we derived in this work with previous pieces of information (the presence or not of outflow, $T_{\text{bol}}$), we can reassess the nature of our source sample more accurately. Even though historically Class 0 and Class I protostars have been classified based on single-dish observations, we propose here to determine the source's evolutionary stage, using the criteria summarised in Table \ref{tab:criteria}. We note that two of our sample sources are close multiple systems (CSO33-b and MMS2-a). Since our 246.2 GHz data does not allow us to disentangle those multiple systems, we   analyse them as single systems. The evolutionary stage of each source is presented in Table \ref{tab:fit_results}.

\begin{table*}[]
    \centering
    \caption{Proposed criteria for defining the evolutionary status of our sample sources based on the derived dust emissivity spectral index, $\beta$; the dust and bolometric temperatures,$T_{\text{d}}$ and $T_{\text{bol}}$; the presence or not of an outflow; and the presence or not of a compact radio continuum source. The range of the dust emissivity spectral index and temperature are based on the discussion in Sect. \ref{sec:review}.}
    \label{tab:criteria}
    \begin{tabular}{ccccc}
    \hline \hline
        \multirow{2}{*}{Evolutionary status} & \multirow{2}{*}{$\beta$ }& $T_{\text{d}}$ / $T_{\text{bol}}$& Outflow & Compact radio continuum\\
        & & K&(Y/N) & (Y/(Y)/N)\\
        \hline
        PSC &1$-$2&$T_{\text{d}}\leq 10$&N&N \\
        Class 0 & $\leq 1.5$&$T_{\text{bol}}<70$&Y&Y  \\
        Class  I& $\leq 1$&$T_{\text{bol}}\geq70$&Y&Y\\
        More evolved ($\geq$Class II) &0$-$1&$T_{\text{d}}\in [10-100]$&Y&(Y)  \\
         \hline
    \end{tabular}
\end{table*}

\subsubsection{Sources whose previous classification is confirmed}
Overall, we could confirm the evolutionary stage of ten sources,  six confirmed Class I protostars and four confirmed Class I protostars.

\paragraph{\textbf{Class 0:}}
The presence of an outflow and of a compact radio continuum emission are two characteristics of the Class 0 and I sources (e.g. \citealt{andre93,anglada1996, bontemps96,barsony1998, andre1997}). FIR1a-b, MMS5, SIMBA-a, and FIR6c-a are sources for which the two properties are clearly observed as an outflow was   detected towards these sources in previous studies \citep[e.g.][]{aso2000, williams2003, takahashi2008, tobin2016b, tanabe2019, feddersen2020, nagy2020}. A large outflow was also   detected towards the MMS9 and CSO33 systems, but the angular resolution does not allow us to determine which of the components is actually the driving source \citep[e.g.][]{williams2003, takahashi2008, tanabe2019, feddersen2020}. 

The dust emissivity spectral indexes are consistent with a Class 0 classification, although in the cases of FIR6c-a and SIMBA-a the index could be consistent with a Class I classification. However, for those sources, and for MMS5 and MMS9-a, the derived range of dust temperatures is surprisingly different from the previously derived bolometric temperature. This indicates that these objects have a large optically thick envelope (see Sect. \ref{subsec:results_temps}), testimony of their young age. They have one of the widest ranges of derived (envelope + disk) masses and temperatures. Furthermore, the derived dust optical depths derived in this work are high enough at low temperature to support this hypothesis (see Table \ref{tab:fit_results}). The derived dust temperature of at least 80 K and the derived range of values of dust emissivity spectral index of MMS9-a is consistent with a Class 0 classification. In the case of FIR1a-b and CSO33-b, the dust temperature is not constrained enough to use this criteria; however, a high percentage of free-free emission is detected, which indicates that the two sources are quite young. In summary,  CSO33-b, FIR6c-a, FIR1a-b, MMS9-a, MMS5, and SIMBA-a are very likely Class 0 sources. 

\paragraph{\textbf{Class I/Flat:}}
Four sources (FIR2, MMS2-a, MMS2-b, and CSO3-b) were previously classified as Class I/Flat spectrum sources, and their status is confirmed with this work. However, we note that this work does not allow us to differentiate between a Class I and a Flat spectrum classification. Therefore, we do not make the distinction between the two classifications, and we  use  the term `{Class I}' only to designate either of the two possibilities.

The derived dust temperature for  MMS2-a and MMS2-b are at least 120 and 80 K respectively, high enough to be consistent with a Class I classification. In the case of CSO3-b and FIR2 the bolometric temperatures of 123 and 108 K, respectively, consistent with the derived dust temperature range, are the reason why these sources are classified as a Class I sources. Then, the derived dust emissivity spectral index for MMS2-a and MMS2-b is $\leq 1$, coherent with this classification. For FIR2 and CSO3-b the dust emissivity spectral index could not be constrained enough. 

Free-free excess has been detected only towards FIR2, which could indicate that the source is a bit less evolved than MMS2-a and MMS2-b. An outflow has been detected towards the MMS2 system (e.g. \citealt{aso2000, williams2003, takahashi2008, tanabe2019, feddersen2020}), but we need high angular resolution studies to understand which of the two sources is the driving source, and thus which of the two sources is younger.  Finally, in the case of CSO3-b, we can see that the 32.9 GHz emission is quite weak, but we cannot determine whether free-free emission is present. Additionally, no sign of outflow features has been detected towards CSO3-b (e.g. \citealt{williams2003, takahashi2008}) and the derived radiative temperature is significantly higher than the other protostars. This could indicate that CSO3-b is more evolved than the three other sources. In summary,  FIR2, MMS2-a, MMS2-b, and CSO3-b are very likely Class I/Flat sources. 

\subsubsection{Sources whose previous classification is modified}

\paragraph{\textbf{FIR1a-a:}}
FIR1a-a was previously   classified as a pre-stellar core \citep{tobin2015, tobin2016b, kainunlainen2017} because no outflow is detected towards this source (e.g. \citealt{tobin2016b, nagy2020}). However, the presence of compact radio emission at 32.9 GHz is incompatible with this classification \citep{bontemps96, yun96}. The dust temperature derived is less than 40 K suggesting a cold source, but the size of the source having been likely overestimated (see Sect. \ref{subsec:spe_cases}), the temperature and dust optical thickness could be underestimated. The range of dust emissivity spectral index values is  compatible with either a Class 0 or I classification, although it is more consistent with a Class I classification ($\beta \in [0.3;0.4]$). No free-free emission is detected at 32.9 GHz towards the source which could support this classification. In summary, FIR1a-a could be either a Class 0 or I protostar, but further investigations are needed to decide on one class over the other.

\paragraph{\textbf{CSO33-a,  MMS9-b, and MMS9-c:}}
CSO33-a, and MMS9-b and MMS9-c are part of the Class 0 systems CSO33 and MMS9, respectively. A compact centimetre emission is also detected towards CSO33-a, MMS9-b, and MMS9-c, but whether they drive an outflow is unknown. Outflows have been detected towards CSO33 and MMS9, but the angular resolution of the observations was insufficient to determine which of the sources actually drives them (e.g. \citealt{williams2003, takahashi2008, tanabe2019, feddersen2020}). The dust emissivity spectral indexes of these sources are very different, with $\beta$ $\leq 0.7$ for CSO33-a, $\beta =0$ for MMS9-b, and $\beta \in$ [0.3;1.8] for MMS9-c.  The spectral indexes are consistent with both  Class 0  and I, except for MMS9-b for which the spectral index is consistent  with a Class I or even Class II classification. However, as seen on the map of the MMS9 field, (Fig. \ref{fig:cont_OMC23}), extended dust is still present around MMS9-b, which therefore indicates that a Class II classification is not suitable. 

Whilst MMS9-b has a derived dust temperature  of at least 50 K, which is consistent with a Class 0 source classification, the dust temperature for all the other sources is not sufficiently constrained. For CSO33-a, a recent study from \citet{sheehan2020} shows evidence of a disk embedded in a massive envelope, which could indicate that the source is quite young and would favour a Class 0 classification. This is consistent with the fact that the derived (envelope+disk) mass for this source is the highest in the source sample. On the other hand, for MMS9-b and MMS9-c, the relatively narrow (envelope+disk) mass range derived would favour a Class I classification. Free-free emission is detected only towards  MMS9-c, but in order to determine if the source could be younger than the other two sources, we need more information on the origin of the free-free emission. In summary, further investigations on the gas temperature and the presence or absence of outflows will help us to decide between  Class 0 or I  for CSO33-a, MMS9-b, and MMS9-c.

\paragraph{\textbf{CSO33-c:}}
CSO33-c is part of the Class 0 system CSO33. However, several parameters show that the source is more evolved than a Class 0 or Class I protostar. First, the source is extremely compact, is rather isolated, and resembles a disk. The dust emissivity spectral index is about 0, which is consistent with what is found for protoplanetary disks of Class II sources (e.g. \citealt{BS91, ricci2010a, ricci2012, perez2012,tazzari2020}).  In summary, CSO33-c is very likely a Class II source.

\paragraph{\textbf{MMS9-d:}}
Finally, MMS9-d is the most puzzling source. This source  is cold ($T_{\text{d}} \leq 20$ K) and closely resembles  CSO33-c in terms of geometry. The dust emissivity spectral index range is not constrained, which does not help us to understand the status of this source. Thus, it could be a Class II protostar with a disk seen edge-on as it is cold. However, our 246.2 GHz ALMA data include a C$^{18}$O line. We thus examined the C$^{18}$O emission towards both CSO33-c and MMS9-d in order to be sure that those objects are associated with the filament. A clear C$^{18}$O emission line is found at the position of CSO33-c, but it is not clear whether the C$^{18}$O emission is associated with MMS9-d (see Fig. \ref{fig:c18o}). In addition,  the source does not show emission at 32.9 GHz. In summary, the source is either evolved ($>$ Class II) or a background object (perhaps a galaxy).

\begin{figure}[ht]
    \centering
    \includegraphics[scale=0.6]{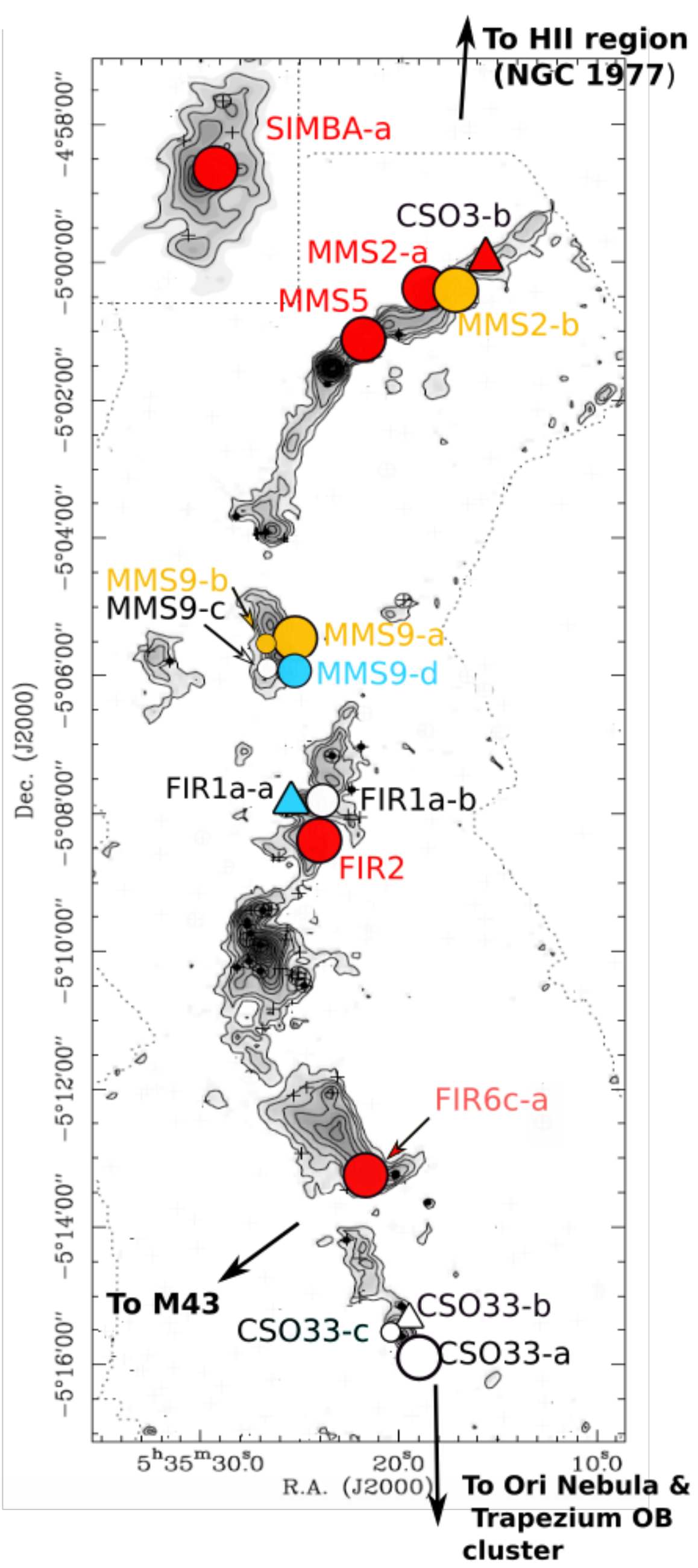}
    \caption{Map of the continuum at 246.2 GHz towards the OMC-2/3 filament, adapted from \citet{chini1997} and \citet{nielbock2003}. The upper left insert (square grid) is taken from the SIMBA map of \citet{nielbock2003}. The dotted line indicates the limits of the continuum map at 1.3mm. Contours rise linearly from a 3$\sigma$ rms noise level, the rms noise of the map being 25 mJy/beam \citep{chini1997}. The sources from our sample are shown as  coloured circles and triangles, colour-coded   depending on the ranges of dust temperature derived in this work: blue means cold ($T_{d} < 50$ K), yellow means warm ($T_{d}\geq 50$ K), red means hot ($T_{d}\geq 90$ K), and white means that the temperature is not constrained enough. The shapes of the symbols  are organised from $M_{\text{env+disk}}$ derived values: the smallest circles for sources with a range of values $< 0.01$\msol, the largest circles for sources with the highest range of values ($\geq0.01$\msol), medium circles for sources with an intermediate range of values, and triangles for lower limits. The three black arrows pointing outside of the filament show the direction to the  different HII regions surrounding OMC-2/3. Some black crosses and filled dots are present throughout the figure, and represent some of the 2MASS sources and MIR sources  from TIMMI 2 observed by \citet{nielbock2003}.}
    \label{fig:omc-filament}
\end{figure}

\subsection{Environment influence}

One of the goals of this study is to understand whether the environment plays a role in the dust properties of the OMC-2/3 filament protostars. As said before, the filament is bounded by several HII regions which illuminate the region. In a previous work it was shown that the large-scale ($\leq 10^4$ au) chemical composition of the gas around the present source sample is largely governed by the UV photons emitted by the massive stars powering the HII regions \citep{bouvier2020}. In this work our aim was to understand whether this highly illuminated environment influences the small-scale (protostellar cores) properties as well. To this end, we studied whether all parameters previously derived ($\beta$, $T_{\text{d}}$, $M_{\text{env+disk}}$, free-free emission) depend on the position of the protostars in the filament.

\begin{figure}[ht]
    \centering
    \includegraphics[width=\linewidth]{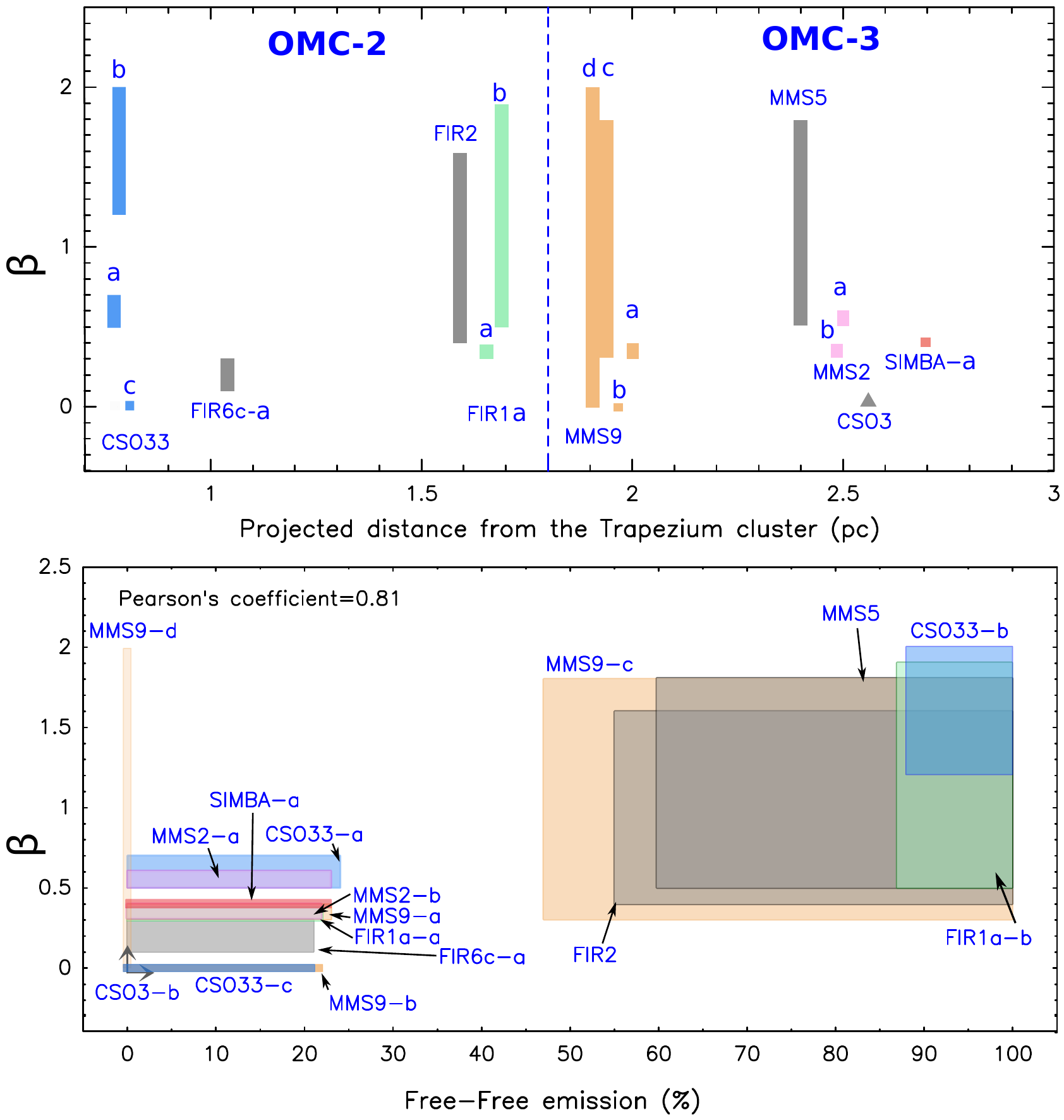}
    \caption{Main results obtained from the dust parameters derivation of the sources. \textit{Top:} Spectral emissivity index, $\beta$, for each source as a function of the position in the OMC-2/3 filament. The dashed blue line indicates the separation between the OMC-2 and OMC-3 clouds. \textit{Bottom:} Spectral emissivity index, $\beta$, as a function of the percentage of free-free emission derived for each source. In both panels, the grey filled rectangles and arrows represent the values of the parameters ($\beta$, free-free emission) associated with the bolometric and/or radiative temperatures of the corresponding source. For clarity, a colour is associated with each   system (blue for CSO33, green for FIR1a, orange for MMS9, and pink for MMS2). The possible values for these sources are represented with their corresponding coloured area.}
    \label{fig:params_fil}
\end{figure}

First, most of the hot sources are located in the north of the filament, in OMC-3, whilst less warm and cold sources tend to be located in the middle of the filament (towards the MMS9 and FIR1a regions), as seen in Fig. \ref{fig:omc-filament}. The  distance between the NGC1977 HII region and SIMBA-a is about the same as the distance between CSO33-a and the Trapezium OB cluster to the south ($\sim 1$pc in both cases). M43 is closer than the other  two  HII regions ($\sim 0.3 $ pc) from the protostars. Therefore, protostars at the edge of the filament could be warmer due to the nearby HII regions. However, we do not have enough information on the temperature of the southern sources (CSO33-a, CSO33-b, and CSO33-c) in order to draw final conclusions on whether the nearby HII regions play a role in the dust temperature of the sources. Moreover, from Fig. \ref{fig:omc-filament}, we do not find any correlation between the (envelope+disk) mass of the sources and their position in the filament. These results show  a different trend than that found in other systems, where the  disk and dust properties of the sources are influenced by nearby sources of UV illumination \citep[e.g.][]{eisner2018,haworth2021, brand2021}. This discrepancy is very likely due to the fact that  our sources are located further away from the UV illumination source than in those studies, and that they are more shielded from the UV illumination as they are still embedded in the molecular cloud. 

Second, we found no correlation between the dust emissivity spectral index, $\beta$, and the positions of the sources in the OMC-2/3 filament, as seen in Fig. \ref{fig:params_fil}, implying that the environment of the OMC-2/3 filament does not affect the evolutionary status of the sources. This would tend to suggest  an almost simultaneous star formation along the OMC-2/3 filament, in agreement with the idea that the latter are structures formed by a relatively fast compression. This result is strengthened by the fact that the more massive protostars ($\geq 0.1$\msol)  seem to be located throughout the filament with no particular trend, as shown in Fig. \ref{fig:omc-filament}.

Third, we found a correlation between $\beta$ and the percentage of free-free emission, with a Pearson correlation coefficient of 0.81 (corresponding to a probability of 66\%), as shown in Fig. \ref{fig:params_fil}. Therefore, objects with higher dust emissivity spectral index seem to show higher free-free emission percentage. However, the sources with higher free-free emission percentage are not necessarily the youngest ones (i.e. Class 0) since, for instance, we find free-free emission towards Class I (FIR2) and not towards SIMBA-a (Class 0). This result is consistent with other studies showing no substantial evidence of strength variation of the free-free emission from Class 0 to I sources \citep{pech2016, tychoniec2018b}.

Evidently, the low statistics and relatively large errors in the parameters of the sources where the dust temperature is not well constrained hamper very strong conclusions on the impact of large-scale UV illumination on low-scale star formation.
Even so, the present work already suggests the potentiality of larger statistical studies towards low-mass star-forming regions to assess whether the presence of external, nearby UV sources has any impact on the forming protostars.


\section{Conclusions}\label{sec:conclusion}
Using new 246.2 GHz ALMA data from the ORANGES project complemented with archival 333 GHz ALMA and 32.9 GHz VLA data from the VANDAM survey \citep{tobin2019a,tobin2019b,tobin2020}, we have constructed the 100 au scale dust SED in the millimetre to centimetre range of 16 protostars located in the OMC-2/3 filament. We have derived several dust parameters, such as the dust temperature, the dust emissivity spectral index, the (envelope+disk) mass of the protostars, and whether free-free emission is present towards the sources. Here are our main conclusions:

\begin{itemize}
    \item From the dust continuum maps, we detected  28  sources, of which 16 could be analysed in this study. Almost all the sources are detected at 333, 246.2, and 32.9 GHz. 
    \item (67 $\pm$ 33)\% of our fields reveal multiple systems, with  average projected separations of at least 1000 au. There is a possible correlation between the number of  components of a multiple  system and the position of the sources in the filament, with a higher number of components in the south of the filament, but this trend is marginal.
    \item From the derived dust temperatures, we could classify the sources in four categories: cold ($T_{\text{d}} < 50$ K), warm ($T_{\text{d}}\geq 50$ K), hot ($T_{\text{d}}>90$ K), and unknown ($T_{\text{d}}$ not constrained). For three sources, \textit{Herschel} likely  probes the large gas envelope surrounding the small-scale protostars, which results in bolometric temperatures lower than the dust temperatures derived in this paper, and indicating that these sources are quite young.
    \item The dust emissivity spectral index of the sources ranges between 0 and 2, with a majority of sources with $\beta < 1$. This range of values coincides with what has been derived in other Class 0 and I sources. Several parameters could lead to a low index, such as grain growth in disks already formed, or the composition of the interstellar grains.
    \item Our results are consistent with the presence of free-free emission in 5 out of 16 sources, implying that they are relatively young sources. A positive correlation between the dust emissivity spectral index and free-free emission is in agreement with the fact that younger objects (large $\beta$) are usually accompanied by free-free emission.
    \item We were able to confirm or correct the evolutionary stage of our source sample, with six Class 0 sources, four Class I sources, one Class II source, and one source that could be either a Class II or a background object. For the remaining four sources, further investigation is needed to decide between a Class 0 or I evolutionary stage. 
    \item The highly illuminated environment of the OMC-2/3 filament does not seem to affect the small-scale structure and dust properties of the OMC-2/3 protostars, although the sources at the edges of the filament could be warmer due to the nearby HII sources. However, we need to better constrain the dust temperature of the sources in order to confirm this result.
\end{itemize}

\begin{acknowledgements}
We deeply thank the referee for their careful reading and their precise and constructive comments that helped to significantly improve the paper. The authors wish to thank F. Motte, F.-X. D\'esert and M. Benisty for their fruitful help on the dust continuum method and useful suggestions and discussions. This project has received funding from the European Research Council (ERC) under the European Union's Horizon 2020 research and innovation programme, for the Project \say{The Dawn of Organic Chemistry} (DOC), grant agreement No 741002. This paper makes use of the following ALMA data: ADS/JAO.ALMA\#2016.1.00376.S and ADS/JAO.ALMA\#2015.1.00041.S. ALMA is a partnership of ESO (representing its member states), NSF (USA) and NINS (Japan), together with NRC (Canada), MOST and ASIAA (Taiwan), and KASI (Republic of Korea), in cooperation with the Republic of Chile. The Joint ALMA Observatory is operated by ESO, AUI/NRAO and NAOJ.

\end{acknowledgements}

\bibliographystyle{aa}
\bibliography{refs}

\appendix

\section{Large fields maps}
\begin{figure*}
    \centering
    \includegraphics[width=0.65\linewidth]{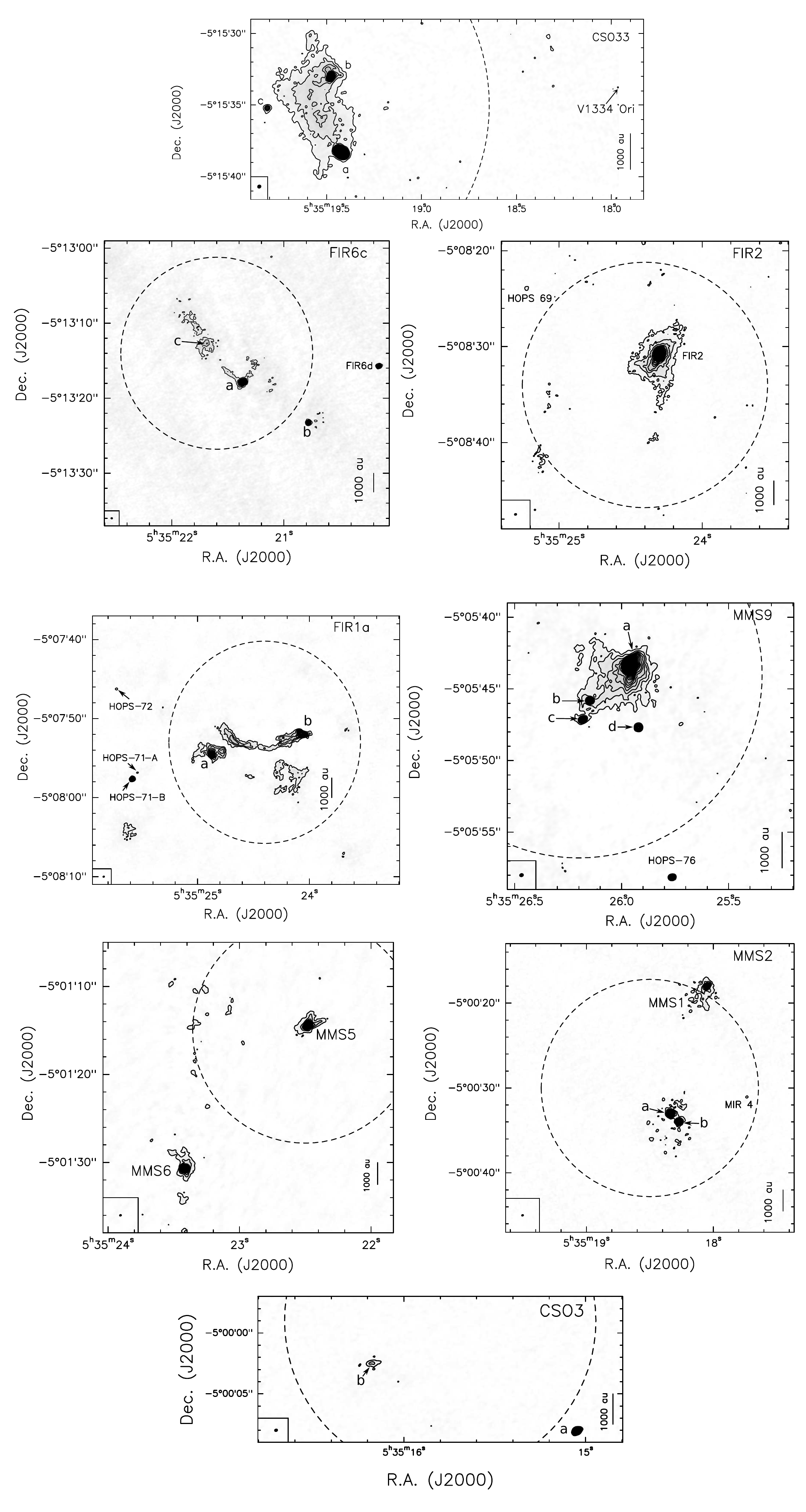}
    \caption{246.2 GHz continuum maps of some of the sample sources with a large field of view to show the detection of the sources located outside the primary beam. Contour levels start at 5$\sigma$ except for CSO33, FIR2 and CSO3 for which contour levels start at 3$\sigma$. Level steps are 5$\sigma$ for FIR6c, FIR2, MMS9, MMS2 and CSO3, 7$\sigma$ for CSO33 and FIR1a and 10$\sigma$ for MMS5. The primary beam is indicated by the dashed circles  and the beam size is represented at the bottom left of the plots. These maps are not corrected for the primary beam. }
    \label{fig:Maps_large}
\end{figure*}

 Table \ref{tab:all_sources} contains the list of all detected sources from the nine fields. The maps of the 246.2 GHz ALMA data showing all the detected sources in each field are shown in Fig. \ref{fig:Maps_large}. The field of SIMBA-a is not shown as SIMBA-a is the only source detected within its field.

\begin{table}[ht]
 \caption{All detected sources in the nine observed fields, with their coordinates and  equivalent name in the HOPS nomenclature. }
    \label{tab:all_sources}
    \centering
    \resizebox{\linewidth}{!}{
    \begin{tabular}{clccc}
    \hline\hline
    Field &Source&R.A. (J2000) & Dec.(J2000) &HOPS name \tablefootmark{a} \\ 
    \hline
    \multirow{3}{*}{CSO33}&CSO33-a &05:35:19.41 &$-$05:15:38.41&HOPS-56-B  \\
    &CSO33-b &05:35:19.48 &$-$05:15:33.08&HOPS-56-A-A/B/C  \\
    &CSO33-c & 05:35:19.81&$-$05:15:35.22&V2358 Ori  \\
    &V1334 Ori & 05:35:17.92 & $-$05:15:32.84&HOPS-56-D\\
    \hline
    \multirow{4}{*}{FIR6c}&FIR6c-a &05:35:21.36&$-$05:13:17.85  &HOPS-409 \\
    &FIIR6c-c& 05:35:21.69 & $-$05:13:12.69 & ... \\
    &FIR6c-b& 05:35:20.78& $-$05:13:23.24  & ...  \\
    &FIR6d &05:35:20.15 &$-$05:13:15.71 &HOPS-59-A  \\
    \hline
    \multirow{2}{*}{FIR2}&FIR2& 05:35:24.30& $-$05:08:30.74 &HOPS-68  \\
    &MIR18 & 05:35:25.23&$-$05:08:23.90 & HOPS-69 \\
    \hline
    \multirow{5}{*}{FIR1-a}&FIR1a-a &05:35:24.87 &$-$05:07:54.63 &HOPS-394-B \\
    &FIR1a-b &05:35:24.05 &$-$05:07:52.07 &HOPS-394-A  \\
    &MIR17-A & 05:35:25.58 & $-$05:07:57.66&HOPS-71-a \\
    &MIR17-B& 05:35:25.54  &$-$05:07:56.83& HOPS-71-b \\
    &MIR16 &05:35:25.72  &$-$05:07:46.25& HOPS-72 \\
    \hline
     \multirow{5}{*}{MMS9}&MMS9-a &05:35:25.97 &$-$05:05:43.34 &HOPS-78-A \\
    &MMS9-b &05:35:26.15 &$-$05:05:45.80 &HOPS-78-B\\
    &MMS9-c &05:35:26.18&$-$05:05:47.14  & HOPS-78-C \\
    &MMS9-d & 05:35:25.92&$-$05:05:47.70 & HOPS-78-D \\
    &MIR12 & 05:35:25.76&$-$05:05:58.15 & HOPS-76  \\
    \hline
     \multirow{2}{*}{MMS5}&MMS6 &05:35:23.42 & $-$05:01:30.53 & HOPS-87  \\
    &MMS5 &05:35:22.47 &$-$05:01:14.34 & HOPS-88\\
    \hline
     \multirow{5}{*}{MMS2}&MMS2-a &05:35:18.34&$-$05:00:32.96  & HOPS-92-A-A/B \\
    &MMS2-b & 05:35:18.27& $-$05:00:33.95 &HOPS-92-B \\
    &MIR4 & 05:35:17.74  &$-$05:00:31.06 & ... \\
    &MMS1 &05:35:18.05 &$-$05:00:18.00 & J05351805-050017.98\\
    \hline
     \multirow{2}{*}{CSO3}&CSO3-a &05:35:15.05&$-$05:00:08.06  & HOPS-93\\
    &CSO3-b &05:35:16.17 &$-$05:00:02.50  &HOPS-94 \\
    \hline
     \multirow{1}{*}{SIMBA-a}&SIMBA-a &05:35:29.72 &$-$04:58:48.60&HOPS-96 \\
    \hline
    \end{tabular}}
    \tablefoot{
\tablefoottext{a}{\citealt{fischer2013,furlan2016}}
}

\end{table}

\section{Fit and residuals}\label{appdx:fit_result_ex}
We present here one example of uv fit result towards SIMBA-a. The best fit was obtained for an elliptical Gaussian of size $0.13 \times 0.11''$ with a position angle of -47. After convolving with the beam, we obtained an ellipse of size $0.35\times 0.29''$ that we used to measured the associated flux densities.

\begin{figure*}
    \centering
    \includegraphics[width=0.8\linewidth]{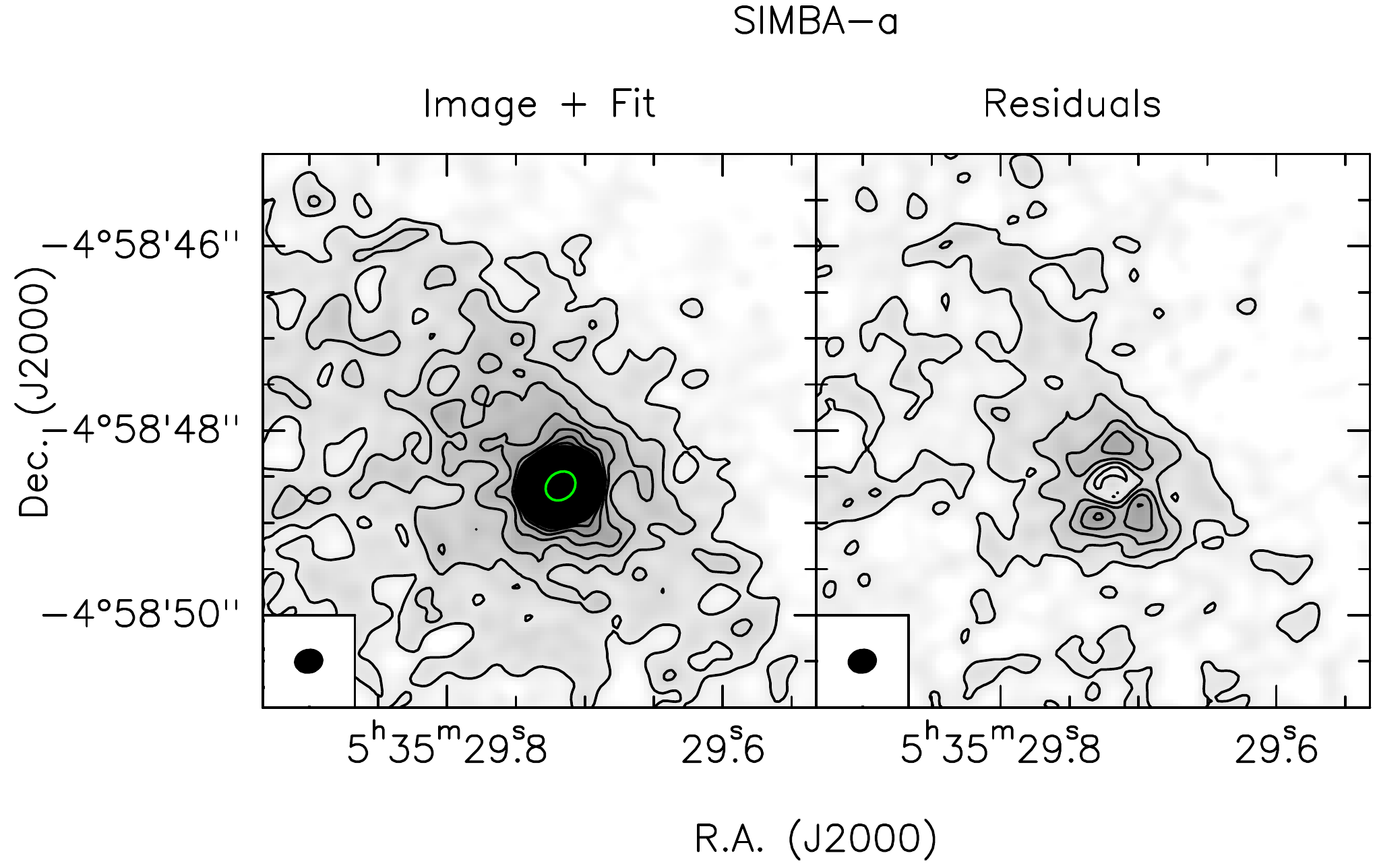}
    \caption{Example of a uv fit for SIMBA-a. \textit{Left:} 246.2 GHz continuum map of SIMBA-a with contours starting from 5$\sigma$ with steps every 5$\sigma$ (See Table \ref{tab:beam_param} for $1\sigma$ values). The source size derived from the fit and convolved by the beam is indicated by the green ellipse. \textit{Right:} Image of the residuals of the uv fit. The synthesised beam is depicted in black in the lower left corner of each panel.}
    \label{fig:c2_res}
\end{figure*}

\section{Dust SED results and uncertainty derivation}

\subsection{Dust parameters}\label{appdx:sed_results}
Plots of the derived free-free emission percentages, the optical depths, the H$_2$ column densities, and the (envelope+disk) masses as a function of the SED temperatures. The results are gathered in Tables \ref{tab:fit_results} and \ref{tab:other_temp}.

\begin{figure*}[ht]
    \centering
    \includegraphics[width=1\linewidth]{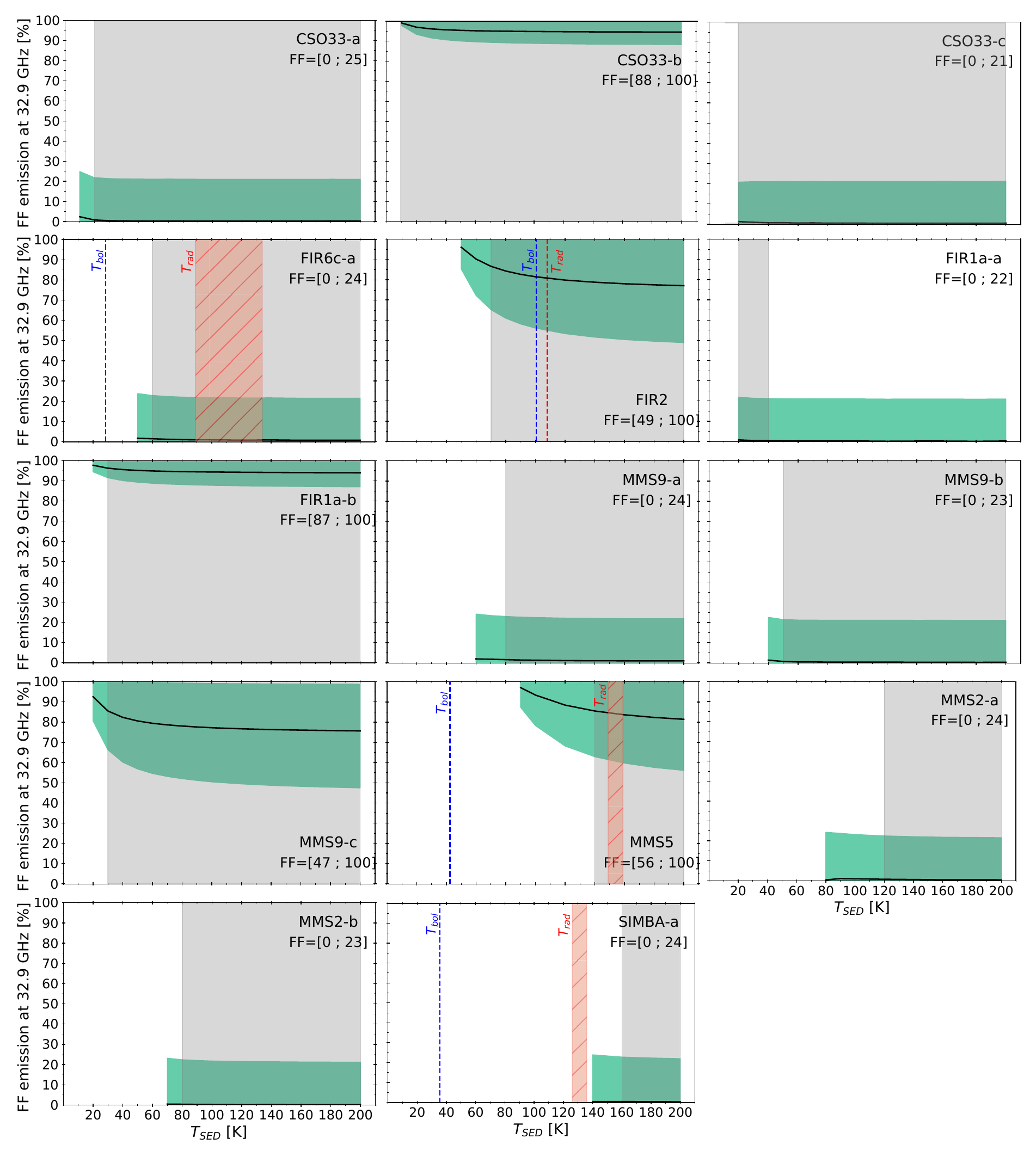}
    \caption{Free-free emission at 32.9 GHz as a function of the SED temperature for each source. The shaded grey area corresponds to the range of temperature for which the $\beta$ values derived are authorised. For each source the minimum and maximum values of the free-free  emission percentage derived over the full temperature range are indicated at the top right of the plots. For single systems, the bolometric temperature, $T_{bol}$, derived from \textit{Herschel} \citep{furlan2016} is indicated by a blue dashed line and the radiative temperature, $T_{rad}$, by a red dashed area or by a red dashed line in the case of FIR2. }
    \label{fig:FF_T_1}
\end{figure*}

\begin{figure*}[ht]
    \centering
    \includegraphics[width=0.9\linewidth]{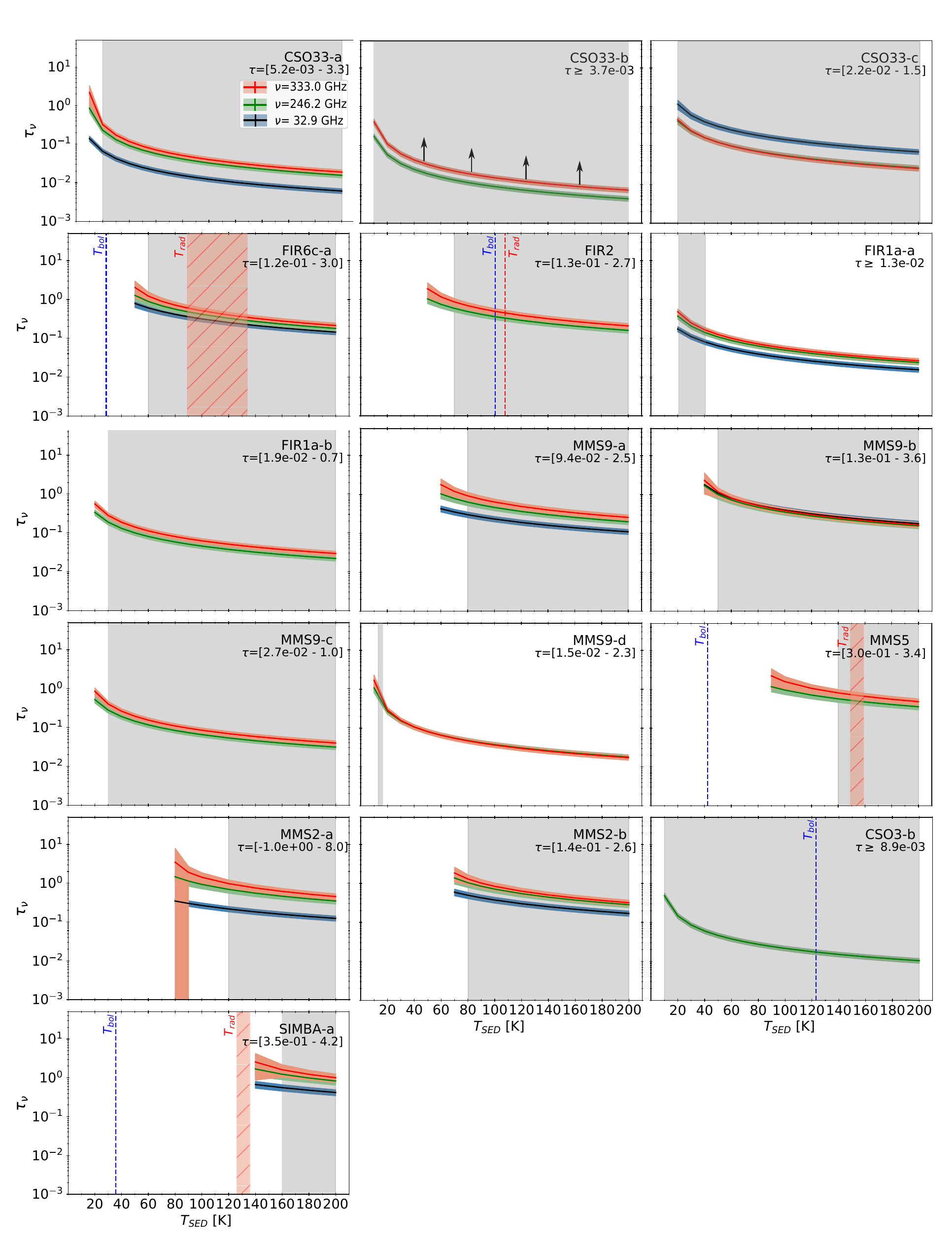}
    \caption{Dust optical depths, $\tau_\nu$, as a function of the SED temperature at 333 GHz (red lines) and at 246.2 GHz (green lines) for each source. The shaded grey area corresponds to the range of temperature for which the $\beta$ values derived are authorised.  The full range of optical depth values derived over the temperature range for each source is indicated on the top right side of the plot. For single systems, the bolometric temperature, $T_{bol}$, derived from \textit{Herschel} \citep{furlan2016} is indicated by a blue dashed line and the radiative temperature, $T_{rad}$, by a red dashed area or by a red dashed line in the case of FIR2. Black arrows indicate that the derived values are lower limits. }
    \label{fig:tau_T_1}
\end{figure*}

\begin{figure*}[ht]
    \centering
    \includegraphics[width=0.9\linewidth]{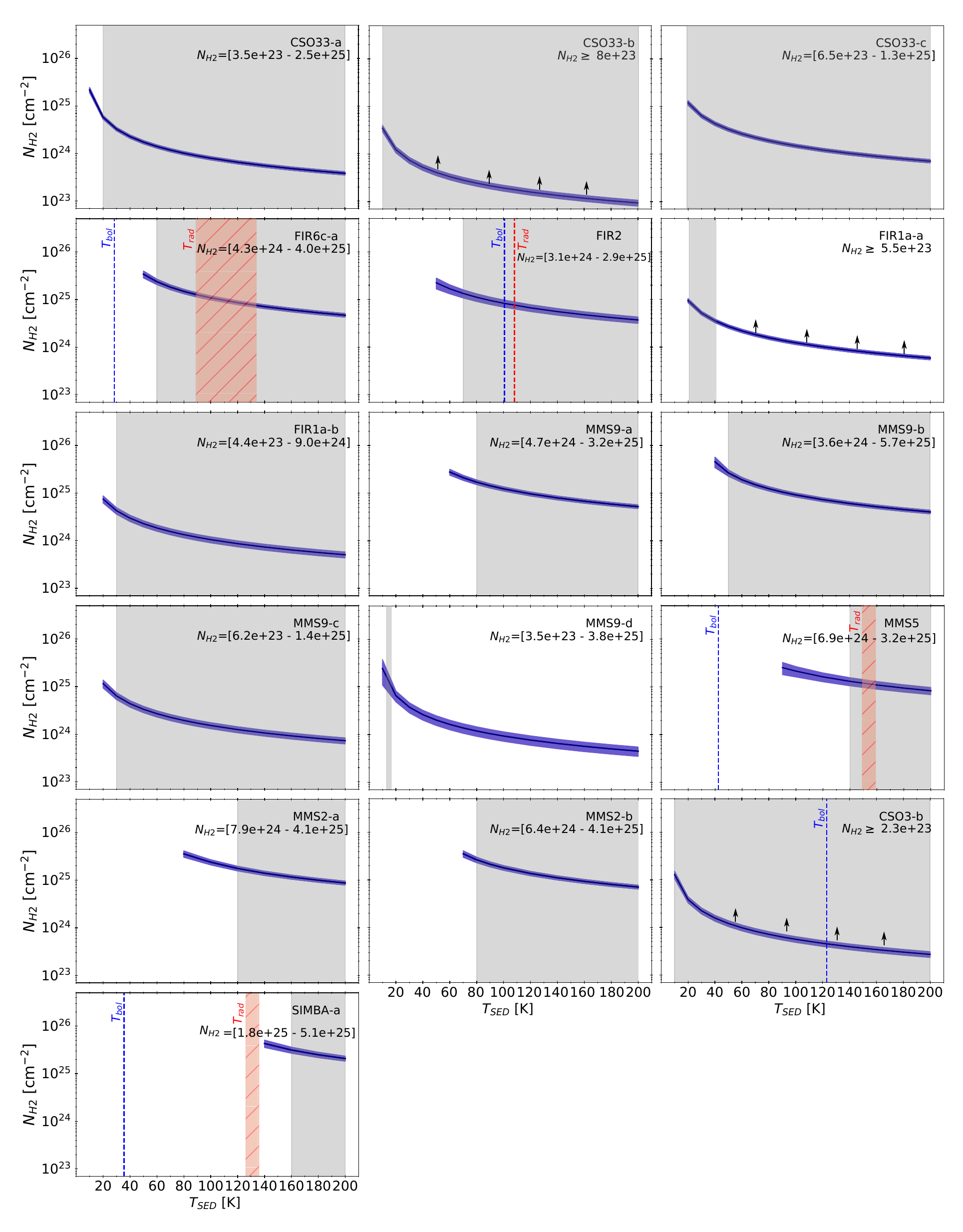}
    \caption{H$_2$ column densities, $N_{H2}$, as a function of the SED temperature for each source. The shaded grey area corresponds to the range of temperature for which the $\beta$ values derived are authorised. For each source the minimum and maximum values of $N_{\text{H2}}$ derived over the full temperature range are indicated at the top right of the plots. For single systems, the bolometric temperature, $T_{bol}$, derived from \textit{Herschel} \citep{furlan2016} is indicated by a blue dashed line and the radiative temperature, $T_{rad}$, by a red dashed area or by a red dashed line in the case of FIR2. Black arrows indicate that the derived values are lower limits.}
    \label{fig:nh2_T_1}
\end{figure*}

\begin{figure*}[ht]
    \centering
    \includegraphics[width=0.9\linewidth]{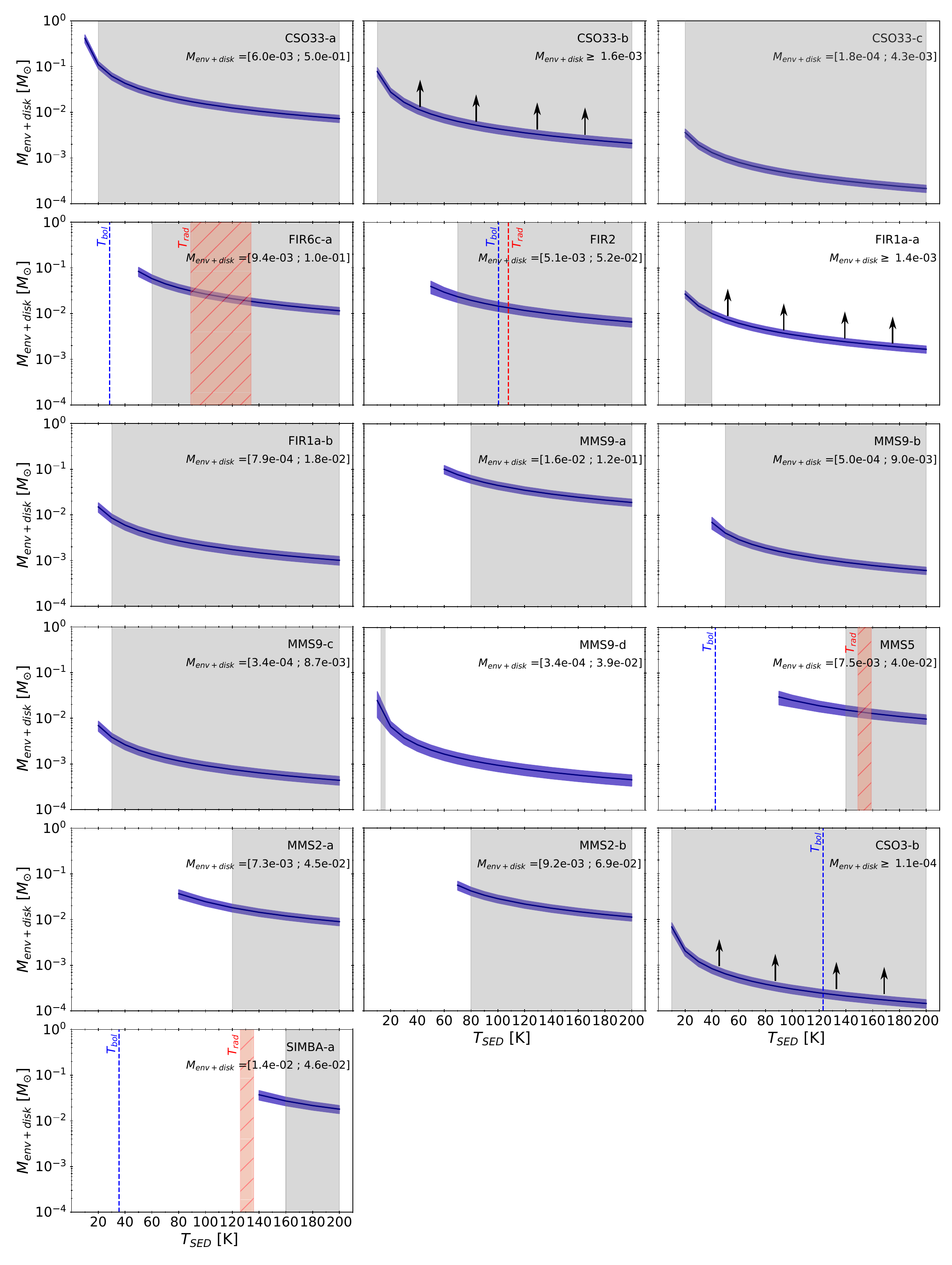}
    \caption{Envelope+disk masses, $M_{\text{env+disk}}$, as a function of the SED temperature for each source. The shaded grey area corresponds to the range of temperature for which the $\beta$ values derived are authorised, which constrains the possible range of $T_{dust}$ values. The range of (envelope+disk) masses derived over the temperature range for each source is indicated on the top side of the plots. For single systems, the bolometric temperature, $T_{bol}$, derived from \textit{Herschel} \citep{furlan2016}, is indicated by a blue dashed line and the radiative temperature, $T_{rad}$, by a red dashed area or by a red dashed line in the case of FIR2. Black arrows indicate that the derived values are lower limits.}
    \label{fig:mass_T_1}
\end{figure*}

\begin{table*}[ht]
    \centering
     \caption{Results of the dust parameter extraction using the SED method} for each source. The range of SED temperature derived for each source follows the constraints given by theoretical and observational works on $\beta$ (see Sect. \ref{sec:review}). First, the linear source sizes are reported in the second column. Then, the SED temperature range is reported in Col. 3. The dust emissivity indexes, between 333 GHz and 246.2 GHz ($\beta_{333-246.2}$), 333 GHz and 32.9 GHz ($\beta_{333-32.9}$), and 246.2 GHz and 32.9 GHz ($\beta_{246.2-32.9}$) are reported in Col. 4. The range of dust optical depth values derived at 333 GHz, $\tau_{333}$, 246.2 GHz, $\tau_{246.2}$, and 32.9 GHz, $\tau_{32.9}$, as well as the H$_2$ column densities, $N_{\text{H2-mm}}$, and (envelope + disk) masses, $M_{\text{env+disk}}$, corresponding to the derived range of temperature, are reported in Cols. 5 to 7. The range of free-free emission percentage at 32.9 GHz is indicated in the last column. \label{tab:fit_results}
    \resizebox{0.85\linewidth}{!}{%
    \begin{tabular}{lccccccc}
    \hline \hline
    \multirow{4}{*}{Source}& \multirow{3}{*}{Size}&\multirow{3}{*}{$T_{\text{SED}}$}&$\beta_{333-246.2}$  &$\tau_{333}$&\multirow{3}{*}{$N_{\text{H2-mm}}$}&\multirow{3}{*}{$M_{\text{env+disk}}$}& \multirow{3}{*}{FF}\\
      &&&$\beta_{333-32.9}$&$\tau_{246.2}$&&\\
      &&&$\beta_{246.2-32.9}$&$\tau_{32.9}$&&\\
      &[au $\times$ au]&[K]&&&[$\times 10^{24}$ cm$^{2}$]&[$\times 10^{-2}$\msol]&[\%]\\
      \hline
         \multirow{3}{*}{CSO33-a} &\multirow{3}{*}{310 $\times$ 153}&\multirow{3}{*}{20 $-$ 200}&0.2 $-$ 1.7&0.02 $-$ 0.4&\multirow{3}{*}{0.3 $-$ 25}&\multirow{3}{*}{1 $-$ 13}&\multirow{3}{*}{0 - 24} \\
        &&&0.4 $-$ 0.8&0.01 $-$ 0.3& && \\
        &&&0.4 $-$ 0.7&0.01 $-$ 0.1&&&\\
        \hline
        \multirow{3}{*}{CSO33-b}&\multirow{3}{*}{236 $\times$ 236}&\multirow{3}{*}{10 $-$ 200} &1.2 $-$ 3.4 &$\geq$ 0.01&\multirow{3}{*}{$\geq$ 0.1}&\multirow{3}{*}{$\geq$ 0.2}&\multirow{3}{*}{88 - 100}\\
        &&&0.3 $-$ 0.6&$\geq$ 0.004&&& \\
        &&&0.2 $-$ 0.4&...&&\\
         \hline
        \multirow{3}{*}{CSO33-c} &\multirow{3}{*}{27.5 $\times$ 27.5}&\multirow{3}{*}{20 $-$ 200}&-0.5 $-$ 0.9 &0.02 $-$ 0.6&\multirow{3}{*}{1 $-$ 13} &\multirow{3}{*}{0.02 $-$ 0.5}&\multirow{3}{*}{0 - 21}\\
        &&&-0.5 $-$ -0.3&0.02 $-$ 0.5& &&  \\
        &&&-0.6 $-$ -0.4&0.1 $-$ 1.5&& \\
         \hline
        \multirow{3}{*}{FIR6c-a} &\multirow{3}{*}{122 $\times$ 51}&\multirow{3}{*}{60 $-$ 200}&0.1 $-$ 2.9 &0.2 $-$ 1.6 & \multirow{3}{*}{4 $-$ 27} &\multirow{3}{*}{1 $-$ 7}&\multirow{3}{*}{0 - 23}\\
        &&&0.1 $-$ 0.6&0.2 $-$ 1.1&& & \\
        &&&0.0 $-$ 0.3&0.1 $-$ 0.7&&&\\
         \hline
        \multirow{3}{*}{FIR2} &\multirow{3}{*}{67 $\times$ 67}&\multirow{3}{*}{70 $-$ 200}&0.4 $-$ 3.2 &0.2 $-$ 1.1&\multirow{3}{*}{3 $-$ 16}&\multirow{3}{*}{0.5 $-$ 3}& \multirow{3}{*}{49 - 100} \\
        &&&0.1 $-$ 0.6 &0.1 $-$ 0.7&&& \\
        &&&0.0 $-$ 0.3 &...&&\\
         \hline
        \multirow{3}{*}{FIR1a-a} &\multirow{3}{*}{90 $\times$ 79}&\multirow{3}{*}{20 $-$ 40}&-0.1 $-$ 1.4&$\geq$ 0.1& \multirow{3}{*}{$\geq$ 3 } &\multirow{3}{*}{$\geq$ 1}& \multirow{3}{*}{0 - 22} \\
        &&&0.2 $-$ 0.5&$\geq$ 0.1&&&\\
        &&&0.1 $-$ 0.5&$\geq 0.1$&& \\
         \hline
        \multirow{3}{*}{FIR1a-b} &\multirow{3}{*}{98 $\times$ 51}&\multirow{3}{*}{30 $-$ 200}&0.5 $-$ 2.2&0.03 $-$ 0.3& \multirow{3}{*}{0.4 $-$ 5 } &\multirow{3}{*}{0.1 $-$ 1}& \multirow{3}{*}{87 - 100}  \\
        &&&-0.3 $-$ 0.0&0.02 $-$ 0.2&& & \\
        &&&-0.5 $-$ -0.3&...&&\\
         \hline
        \multirow{3}{*}{MMS9-a} &\multirow{3}{*}{173 $\times$ 55}&\multirow{3}{*}{80 $-$ 200}&0.3 $-$ 3.0&0.2 $-$ 1.0&\multirow{3}{*}{3 $-$ 19} &\multirow{3}{*}{2 $-$ 7}&\multirow{3}{*}{0 - 23}  \\
        &&&0.3 $-$ 0.8&0.2 $-$ 0.7&& & \\
        &&&0.2 $-$ 0.5&0.1 $-$ 0.4&&\\
         \hline
        \multirow{3}{*}{MMS9-b} &\multirow{3}{*}{20 $\times$ 20}&\multirow{3}{*}{50 $-$ 200}& -0.3 $-$ 2.7&0.1 $-$ 1.5& \multirow{3}{*}{3 $-$ 31}&\multirow{3}{*}{ 0.05 $-$ 0.5}&\multirow{3}{*}{0 - 22} \\
        &&&-0.1 $-$ 0.3&0.1 $-$ 1.3&&& \\
        &&&-0.1 $-$ 0.2&0.2 $-$ 1.3&&\\
         \hline
        \multirow{3}{*}{MMS9-c} &\multirow{3}{*}{39 $\times$ 39}&\multirow{3}{*}{30 $-$ 200}& 0.3 $-$ 2.3&0.03 $-$ 0.5& \multirow{3}{*}{0.6 $-$ 7 } &\multirow{3}{*}{0.03 $-$ 0.5}&\multirow{3}{*}{47 - 100} \\
        &&&0.1 $-$ 0.5&0.03 $-$ 0.3&&&\\
        &&&0.0 $-$ 0.4&... &&\\
         \hline
        \multirow{3}{*}{MMS9-d} &\multirow{3}{*}{51 $\times$ 51}&\multirow{3}{*}{10 $-$ 20}& -0.5 $-$ 2.6 &0.3 $-$ 2.4 & \multirow{3}{*}{5 $-$ 37}&\multirow{3}{*}{0.5 $-$ 4} &\multirow{3}{*}{...}\\
        &&&...&0.3 $-$ 1.4 &&&\\
        &&&...&...&&&  \\
         \hline
        \multirow{3}{*}{MMS5} &\multirow{3}{*}{59 $\times$ 51}&\multirow{3}{*}{140 $-$ 200}&0.4 $-$ 2.7&0.4 $-$ 1.0& \multirow{3}{*}{6 $-$ 16} &\multirow{3}{*}{1 $-$ 2}& \multirow{3}{*}{56 - 100}\\
        &&&0.1 $-$ 0.5&0.3 $-$ 0.7&&&  \\
        &&&0.0 $-$ 0.3&...&&\\
         \hline
        \multirow{3}{*}{MMS2-a} &\multirow{3}{*}{51 $\times$ 51}&\multirow{3}{*}{120 $-$ 200}&0.3 $-$ 3.0 &0.4 $-$ 1.2& \multirow{3}{*}{7 $-$ 20} &\multirow{3}{*}{1 $-$ 2}& \multirow{3}{*}{0 - 23} \\
        &&&0.5 $-$ 1.0&0.3 $-$ 0.9&&&\\
        &&&0.4 $-$ 0.8&0.1 $-$ 0.3&&\\
         \hline
        \multirow{3}{*}{MMS2-b} &\multirow{3}{*}{51 $\times$ 47}&\multirow{3}{*}{80 $-$ 200}&-0.2  $-$ 2.2&0.3 $-$ 1.7& \multirow{3}{*}{6 $-$ 29} &\multirow{3}{*}{1 $-$ 5}&\multirow{3}{*}{0 - 22} \\
         &&&0.2 $-$ 0.6&0.2 $-$ 1.3&&& \\
        &&&0.2 $-$ 0.5&0.1 $-$ 0.6&&\\
         \hline
        \multirow{3}{*}{CSO3-b} &\multirow{3}{*}{...}&\multirow{3}{*}{10 $-$ 200}&... &...&\multirow{3}{*}{$\geq$ 0.2}\tnote{a}&\multirow{3}{*}{$\geq$ 0.01}& \multirow{3}{*}{...}\\
        &&&...&$\geq$ 0.01 &&& \\
        &&&$\geq -1.4$&...&&&  \\
         \hline
        \multirow{3}{*}{SIMBA-a} &\multirow{3}{*}{51 $\times$ 43}&\multirow{3}{*}{160 $-$ 200}&-0.1  $-$ 3.3&0.7 $-$ 2.2& \multirow{3}{*}{16 $-$ 36}  &\multirow{3}{*}{1 $-$ 3}&\multirow{3}{*}{0 - 23} \\
        &&&0.3 $-$ 0.8&0.6 $-$ 1.6&&& \\
        &&&0.2 $-$ 0.6&0.4 $-$ 0.7&&\\
      \hline
    \end{tabular}%
    }
   
\end{table*}

\begin{table*}[ht!]
    \caption{Bolometric and radiative temperatures, $T_{\text{bol}}$ and $T_{\text{rad}}$, for the four sources of our sample that are single systems, their corresponding dust parameters ($\beta$, $M_{\text{env+disk}}$),  and the free-free emission percentage.}
    \label{tab:other_temp}
    \centering
    \resizebox{0.85\linewidth}{!}{%
    \begin{tabular}{lcccccccc}
    \hline \hline
        \multirow{2}{*}{Source} & $T_{\text{bol}}$&$\beta_{\text{Tbol}}$&$M_{\text{Tbol}}$&FF$_{\text{Tbol}}$&$T_{\text{rad}}$&$\beta_{\text{rad}}$&$M_{\text{Trad}}$&FF$_{\text{Trad}}$ \\
         & [K]&&[$\times 10^{-2}$\msol]&[\%]&[K]&&[$\times 10^{-2}$\msol]&[\%] \\
         \hline
         FIR6c-a&28.4&... &...&...&89 $-$ 134&0.1 $-$ 0.3&1.5 $-$ 4&0 $-$ 21\\
         FIR2 &100.6& 0.4 $-$ 1.6&1.1 $-$ 1.9&56 $-$ 100&108&0.4 $-$ 1.6&1.0 $-$ 1.8&55 $-$ 100\\
         MMS5&42.4&...&...&...&149 $-$ 159&0.5 $-$ 1.8&1.0 $-$ 1.9&60 $-$ 100\\
         CSO3-b&123 &$\geq - 0.7$&$\geq 0.02$&$\geq0$&265$-$324&...&...&...\\
         SIMBA-a&35.6&...&...&...&126 $-$ 136&...&...&...\\
         \hline
    \end{tabular}
    }
\end{table*}

\subsection{Derivation of uncertainties }\label{appdx:uncertainties}
In order to derive the uncertainties of the different quantities, we use error propagation. In general, the error of an arbitrary function \textit{q(x$_1$,...,x$_n$)} with uncertainties ($\delta x_1$,...,$\delta x_n$), is

\begin{equation}
    \Delta q=\sqrt{\left(\frac{\partial q}{\partial x_1}\delta x_1\right)^2+...+\left(\frac{\partial q}{\partial x_n}\delta x_n\right)^2}
.\end{equation}
The formula used to derive the uncertainties of the  dust emissivity spectral index, d$\beta$; dust opacity, d$\tau$; H$_2$ column density, d$N_{\text{H2}}$; and (envelope + disk) mass, d$M_{\text{env+disk}}$  are shown  below: 
\begin{equation}
    \begin{split} 
    \Delta\beta^2=&\left(-\frac{\delta F_{\nu1}}{\text{ln}\left(\frac{\nu_1}{\nu_2}\right)\text{ln}\left(1-\frac{F_{\nu1}}{\Omega_sB_{\nu1}}\right)\left(\Omega_sB_{\nu1}-F_{\nu1}\right)}\right)^2 \\ 
    &+ \left(\frac{\delta F_{\nu2}}{\text{ln}\left(\frac{\nu_1}{\nu_2}\right)\text{ln}\left(1-\frac{F_{\nu2}}{\Omega_sB_{\nu2}}\right)\left(\Omega_sB_{\nu2}-F_{\nu2}\right)}\right)^2 \\ 
   &+\left(\frac{\delta\Omega_s}{\text{ln}(\frac{\nu_1}{\nu_2})}\right)^2\left( \frac{F_{\nu1}}{\Omega_s(B_{\nu1}\Omega_s-F_{\nu1})\text{ln}\left(1-\frac{F_{\nu1}}{B_{\nu1}\Omega_s}\right)}\right.\\
   &\left.-\frac{F_{\nu2}}{\Omega_s(B_{\nu2}\Omega_s-F_{\nu2})\text{ln}\left(1-\frac{F_{\nu2}}{B_{\nu2}\Omega_s}\right)}\right)^2
   \end{split}
   ,\end{equation}

\begin{equation}
    \Delta\tau=\frac{1}{B_{\nu}\Omega_s-F_{\nu}}\sqrt{{\delta F_\nu}^2+\left(\frac{F_\nu}{\Omega_s}\delta\Omega_s\right)^2}
,\end{equation}

\begin{equation}
    \Delta N_{\text{H2}}=\frac{1}{\mu m_HR\kappa_0(\frac{\nu}{\nu_0})^{\beta}}\sqrt{{\delta_{\tau_\nu}}^2+\left(\tau_\nu\text{ln}\left(\frac{\nu}{\nu_0}\right)\delta\beta\right)^2}
,\end{equation}

\begin{equation}
    \Delta M_{env+disk}=\mu m_{\text{H}}d \sqrt{ (d\Omega\delta N_{\text{H2}})^2+(2N_{\text{H2}}\Omega\delta d)^2+(N_{\text{H2}}d\delta \Omega)^2}
.\end{equation}

\subsection{Dust SEDs plots}
We show here the obtained dust SEDs for each of the source samples, corresponding to the derived dust parameters ($\beta, T_{\text{d}}$, and $N_{\text{H2}}$). We added to each plot the measured flux densities as a self-consistency check.

\begin{figure*}
    \centering
    \includegraphics[width=\linewidth]{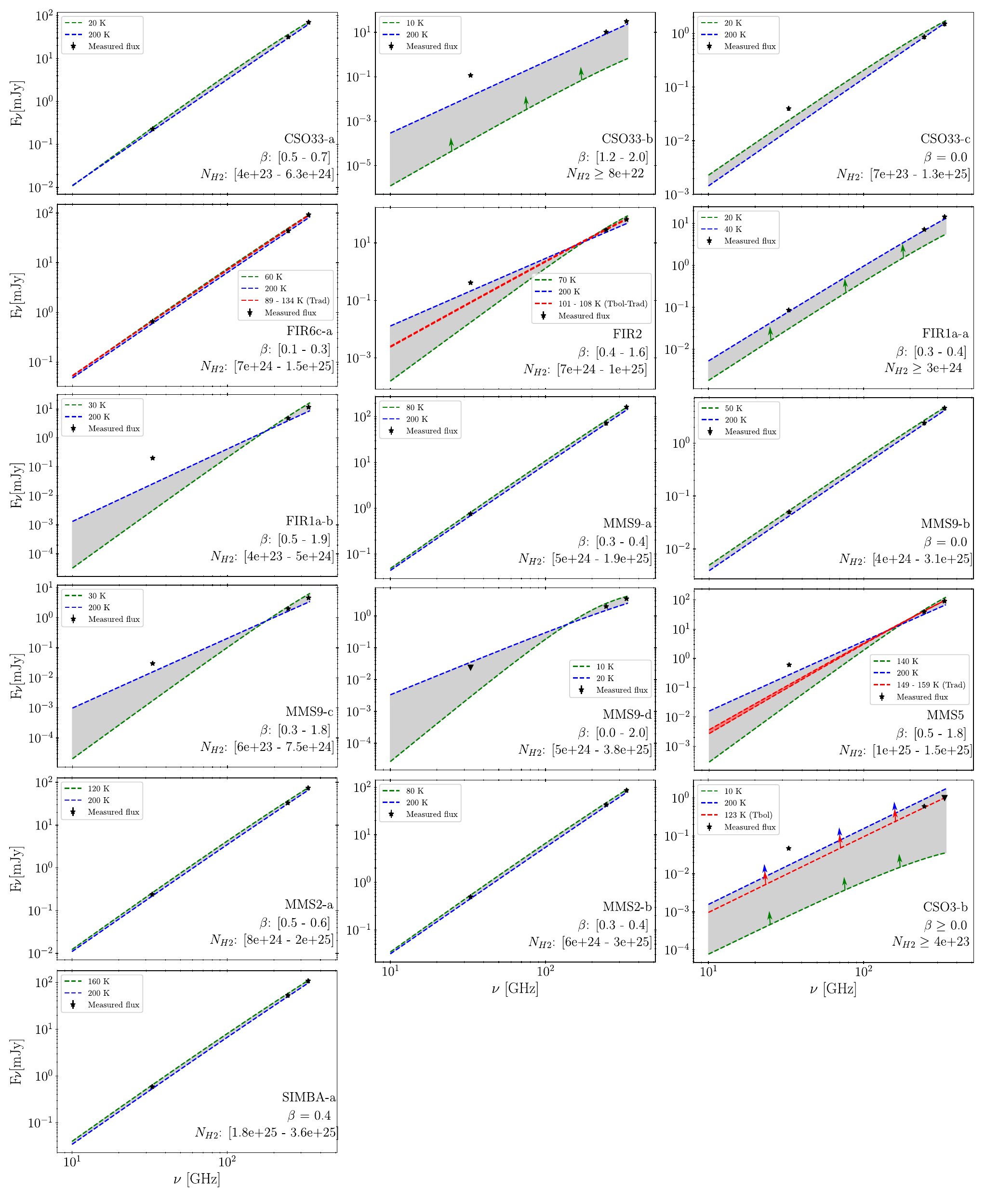}
    \caption{Dust SED obtained for each source from the derived dust temperatures; dust emissivity spectral indexes, $\beta$; and H$_2$ column density, $N_{\text{H2}}$. For each source the dust SED of the two extreme values for the corresponding dust temperature range are indicated by blue and green dashed lines. The possible dust SEDs corresponding to the intermediate values of dust temperatures are represented by the grey shaded area. For FIR6c-a, FIR2, MMS5, and CSO3-b the dust SEDs corresponding to the derived range of $T_{\text{rad}}$ and/or $T_{\text{bol}}$ are also represented by red dashed lines and a red shaded area. Measured flux densities are represented by black stars. Lower limits are represented by coloured arrows.}
    \label{fig:SEDS_plot}
\end{figure*}

\section{C$^{18}$O maps}
We present here the C$^{18}$O ($J=2-1$) maps towards the CSO33 and MMS9 fields.

\begin{figure*}
    \centering
    \includegraphics[width=0.9\linewidth]{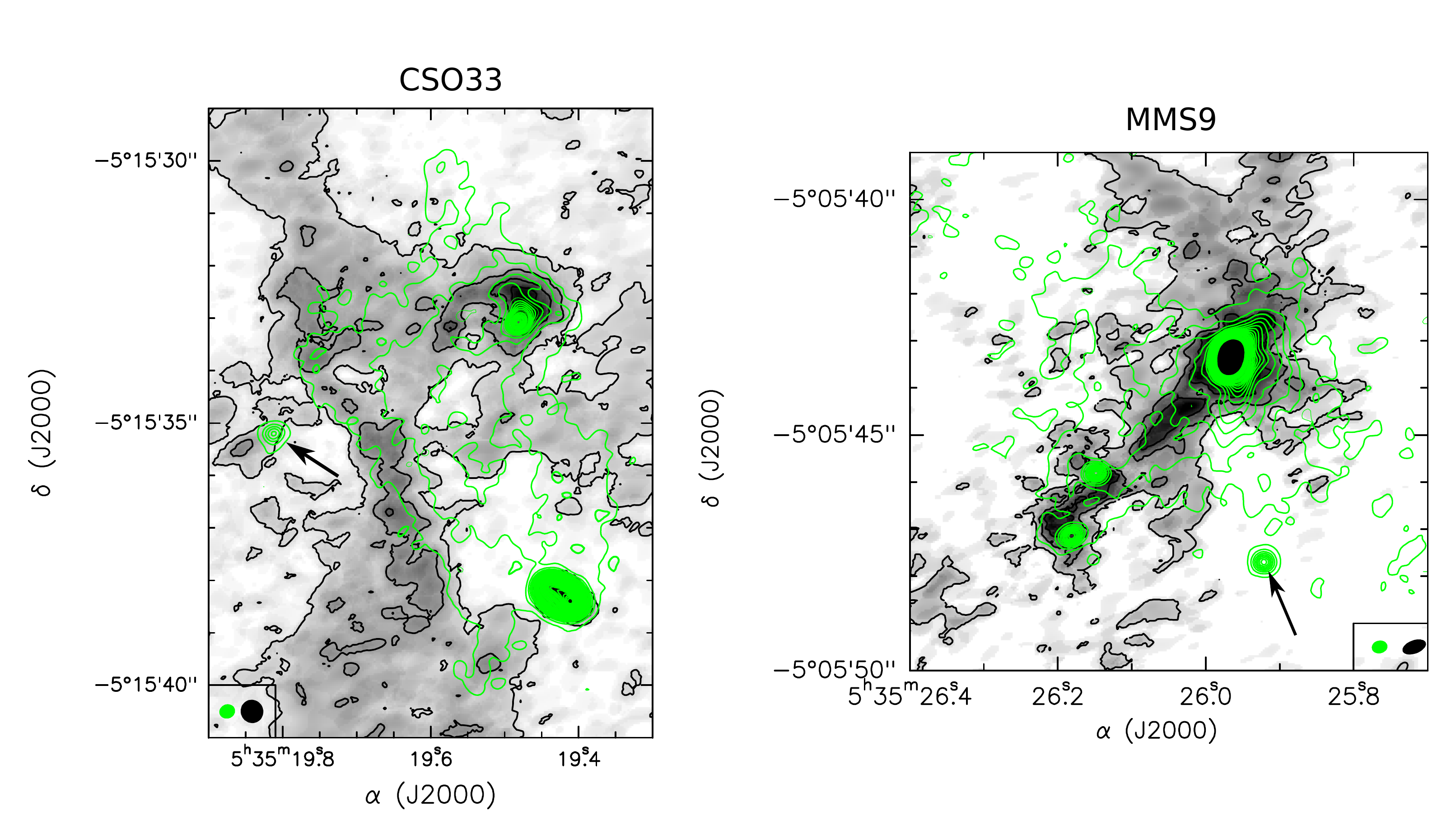}
    \caption{CSO33 and MMS9 maps of C$^{18}$O ($J=2-1$) (in grey shades and black contours) with the 246.2 GHz continuum contours superposed (in green). The C$^{18}$O contours starts at 3$\sigma$ (1$\sigma=4.8$mJy/beam for CSO33 and 1$\sigma=3$ mJy/beam for MMS9) with steps of 5$\sigma$. Continuum contours starts at 5$\sigma$ with steps of 10$\sigma$ (see Table \ref{tab:beam_param} for the 1$\sigma$ values). The complete description for the molecular observations setup will be presented in a forthcoming paper (Bouvier et al. in prep.). Arrows are pointing towards CSO33-c and MMS9-d.  }
    \label{fig:c18o}
\end{figure*}

\end{document}